\documentclass[a4paper,10pt]{IEEEtran}
\usepackage[normalem]{ulem}

\pdfoutput=1
\IEEEoverridecommandlockouts
\makeatletter
\def\endthebibliography{%
	\def\@noitemerr{\@latex@warning{Empty `thebibliography' environment}}%
	\endlist
}
\makeatother

\IEEEoverridecommandlockouts
\usepackage{setspace}
\usepackage{multirow}
\usepackage{lipsum}
\usepackage{amsfonts}
\usepackage{array}
\usepackage{amsmath,cases,amssymb}
\usepackage{amsmath}
\usepackage{amsthm}
\usepackage{graphicx}
\usepackage{cite}
\usepackage{mathtools}
\usepackage{subfigure}
\usepackage{epsfig}
\usepackage{url}
\usepackage{algorithm}
\usepackage{algorithmic}
\usepackage{epstopdf}
\usepackage{color}
\usepackage{makecell}
\usepackage{xcolor}

\usepackage{lipsum}
\usepackage[justification=centering]{caption}

\DeclareGraphicsExtensions{.eps,.pdf}

\newcommand{\bPsi}{\boldsymbol{\Psi}}

\newcommand{\non}{\nonumber}

\newcommand{\bw}{\mathbf{w}}

\newcommand{\bC}{\mathbf{C}}

\newcommand{\bH}{\boldsymbol{H}}

\newcommand{\bB}{\boldsymbol{B}}

\newcommand{\bI}{\mathbf{I}}

\newcommand{\bu}{\mathbf{u}}

\newcommand{\bP}{\mathbf{P}}

\newcommand{\bo}{\boldsymbol{o}}
\newcommand{\bb}{\mathbf{b}}
\newcommand{\ba}{\boldsymbol{a}}
\newcommand{\bs}{\boldsymbol{s}}
\newcommand{\bK}{\mathbf{K}}

\newcommand{\bmu}{\pmb{\mu}}

\newcommand{\Ptt}{\mathtt{P}}
\newcommand{\Ott}{\mathtt{O}}

\newcommand{\Scal}{\mathcal{S}}
\newcommand{\Acal}{\mathcal{A}}

\newcommand{\Qcal}{\mathcal{Q}}
\newcommand{\Kcal}{\mathcal{K}}

\newcommand{\Mcal}{\mathcal{M}}

\newcommand{\Xcal}{\mathcal{X}}
\newcommand{\Ncal}{\mathcal{N}}

\newcommand{\Pcal}{\mathcal{P}}
\newcommand{\Ocal}{\mathcal{O}}
\newcommand{\Tcal}{\mathcal{T}}

\newcommand{\PA}{\mathtt{PA}}
\newcommand{\SA}{\mathtt{SA}}

\makeatletter
\newcommand{\doublewidetilde}[1]{{%
		\mathpalette\double@widetilde{#1}}}
\newcommand{\double@widetilde}[2]{%
	\sbox\z@{$\m@th#1\widetilde{#2}$}%
	\ht\z@=.5\ht\z@
	\widetilde{\box\z@}}
\makeatother

\newtheorem{lemma}{Lemma}

\newtheorem{example}{Example}
\newtheorem{remark}{Remark}

\DeclareMathOperator{\diag}{diag}

\begin{document}
\newcommand{\pp}[1]{\textcolor{red}{#1}}
\newcommand{\phuc}[1]{\textcolor{black}{#1}}
\newcommand{\forest}[1]{\textcolor{orange}{#1}}
\newcommand{\bsa}[1]{\textcolor{teal}{#1}}
	
% \title{Efficient Communication and Control  in Digital Twins under Timing Constraints}
%\title{Towards Efficient Decision Making in Cloud-Based Digital Twins under Timing Constraints}
\title{Timely Communication from Sensors for Wireless Networked Control in Cloud-Based Digital Twins}

\author{ Van-Phuc Bui, \textit{IEEE Student Member}, Shashi Raj Pandey,\textit{ IEEE Member}, Pedro M. de Sant Ana, \\ Beatriz Soret, \textit{Senior Member, IEEE}, Petar Popovski, \textit{IEEE Fellow}\thanks{V.-P Bui, S.R. Pandey, B. Soret and P. Popovski (emails: \{vpb, srp, bsa, petarp\}@es.aau.dk) are all with the Department of Electronic Systems, Aalborg University, Denmark. P. M. de Sant Ana is with the Corporate Research, Robert Bosch GmbH, 71272 Renningen, Germany (email: Pedro.MaiadeSantAna@de.bosch.com). B. Soret is also with the Telecommunications Research Institute (TELMA), University of Malaga. This work was supported in part by the Villum Investigator Grant ``WATER'' from the Velux Foundation, Denmark, and in part by the Horizon Europe SNS project “6G-GOALS” (grant 101139232). An earlier version of this paper was presented in part at the IEEE ICC 2024 \cite{bui2024icc}.} }

\maketitle
\begin{abstract}

We consider a Wireless Networked Control System (WNCS) where sensors provide observations to build a DT model of the underlying system dynamics. The focus is on control, scheduling, and resource allocation for sensory observation to ensure timely delivery to the DT model deployed in the cloud. \phuc{Timely and relevant information, as characterized by optimized data acquisition policy and low latency, are instrumental in ensuring that the DT model can accurately estimate and predict system states. However, optimizing closed-loop  control with DT and acquiring data for efficient state estimation and control computing pose a non-trivial problem given the limited network resources, partial state vector information, and measurement errors encountered at distributed sensing agents.} To address this, we propose the \emph{Age-of-Loop REinforcement learning and Variational Extended Kalman filter with Robust Belief (AoL-REVERB)}, which leverages an uncertainty-control reinforcement learning solution combined with an algorithm based on Value of Information (VoI) for performing optimal control and selecting the most informative sensors to satisfy the prediction accuracy of DT. Numerical results demonstrate that the DT platform can offer satisfactory performance while halving the communication overhead.

\end{abstract}
\begin{IEEEkeywords}
    Digital twin, Reinforcement Learning, Age-of-Loop, Internet of Things, Network Control System.
\end{IEEEkeywords}

%%%%%%%%%%%%%%%%%%%%%%%%%%%%%%%%%%%%%%%%%%%%%%%%
\section{Introduction}\label{sec:intro}
%%%%%%%%%%%%%%%%%%%%%%%%%%%%%%%%%%%%%%%%%%%%%%%%

The advent of Industry 4.0's intelligent manufacturing paradigm mandates the acquisition of substantial real-time data volumes from a diverse array of wireless sensors~\cite{tang2015tracking}. In contrast to conventional simulation tools or optimization methodology, digital twin (DT) models undergo a process of transforming these extensive datasets into predictive models. DTs are seen as a pivotal technological facilitator within wireless cellular systems, adhering to the principles of open networking, characterized by disaggregation and virtualization \cite{li2020ran}.  
The utilization of these models facilitates the emulation of potential control strategies, thereby supporting real-time interactions and decision-making for system operators~\cite{9899718}.

\begin{figure}[t]
    \begin{minipage}{0.5\textwidth}
        \centering
        \includegraphics[width=1\textwidth]{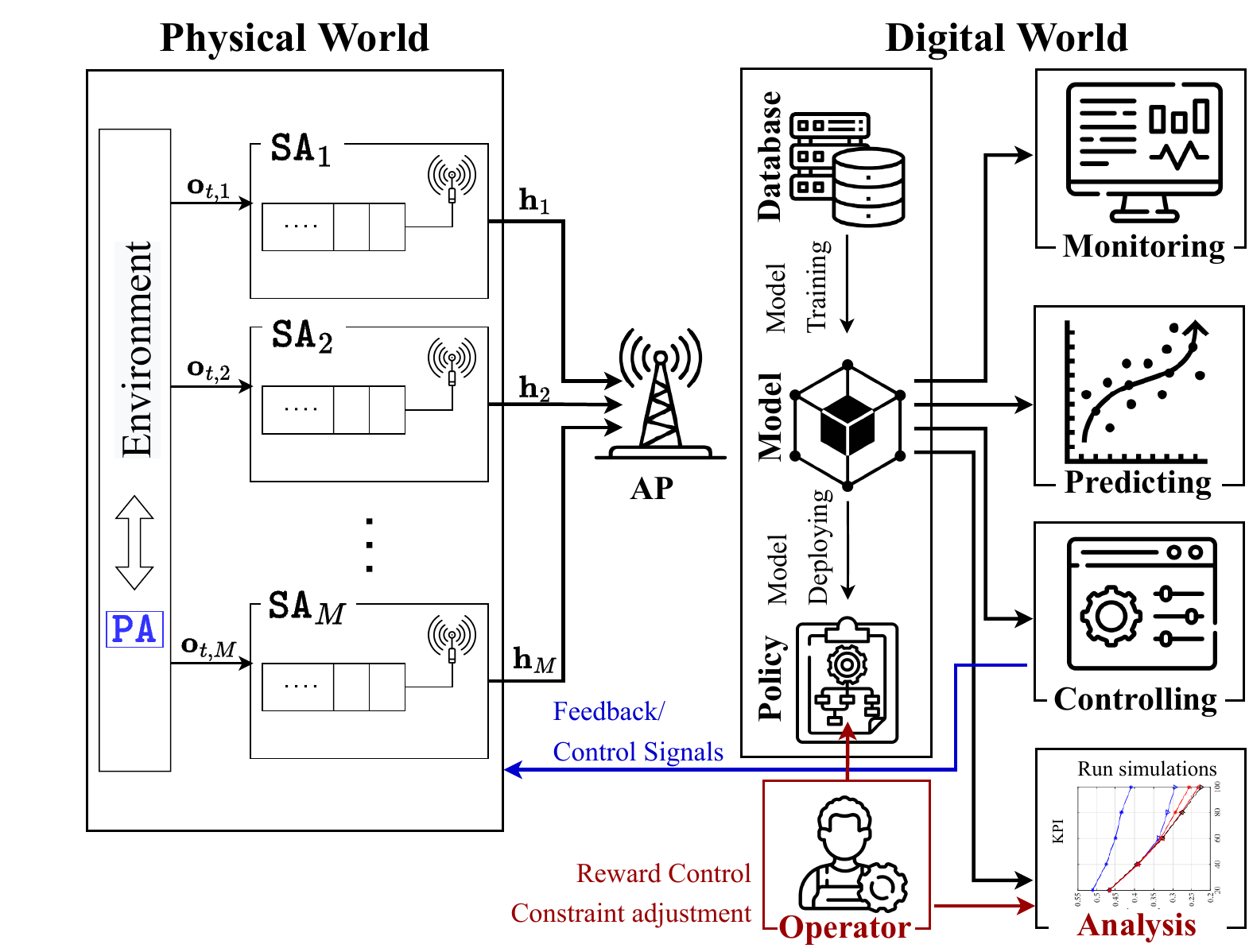} \\ 
        % \caption{heading}
        % \vspace*{-5pt}
        % \caption{The considered architecture with a Digital Twin (DT).}
        \centering {\footnotesize$(a)$}
        \vspace*{5pt}
    \end{minipage}
    % \vspace{10pt}
    \begin{minipage}{0.5\textwidth}
        \centering
        \includegraphics[width=1\textwidth]{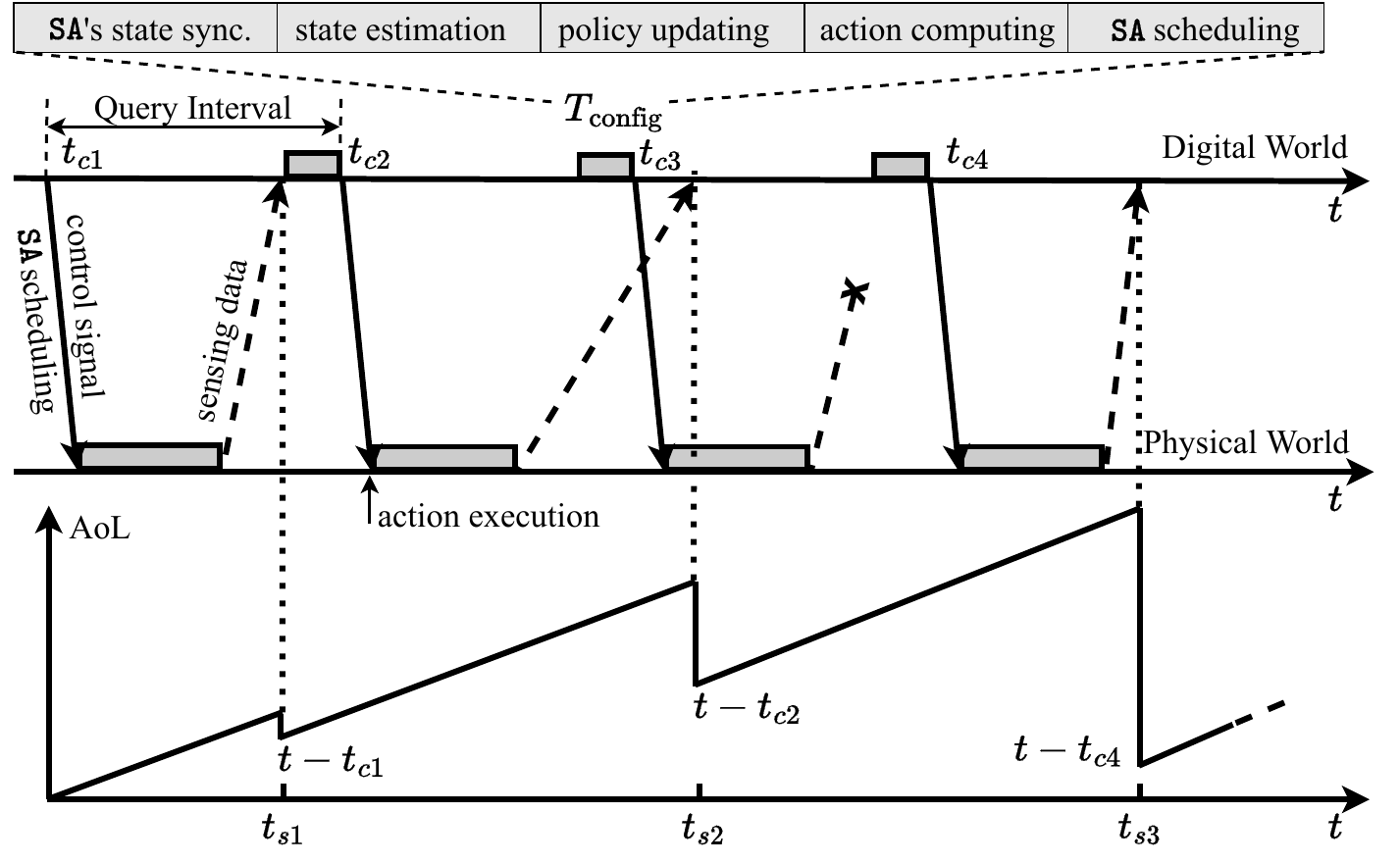} \\ 
        \vspace*{-0pt}
        % \caption{Timing diagram of signals transmitted with corresponding AoL.}
        \centering {\footnotesize$(b)$}
        \vspace*{-0pt}
    \end{minipage}
    \captionsetup{format=plain, justification=justified, width=1\linewidth}
    \caption{\small{The considered architecture with a Digital Twin (DT): $(a)$ The system model; $(b)$ Timing diagram of signals transmitted with corresponding Age-of-Loop (AoL). \phuc{Downlink signals, which transitions from the digital world to the physical world, include the control signals applied at the $\PA$, the scheduling of $\SA$s, and the corresponding communication resource scheduling. The uplinks involve the wireless transmission of observations from the scheduled $\SA$s to the Access Point (AP).}  }}
    \vspace{-20pt}
    \label{fig_system}
\end{figure}

A wireless network control system (WNCS) representing the physical world, alongside a DT is situated either in the cloud, catering to wide-area physical systems, or at the edge, tailored to local physical systems. The network comprises sensor devices and/or central/distributed units integral to a 5G system or beyond \cite{jagannath2022digital}.
The DT is responsible for gathering data from the physical world, either periodically or adaptively, which is subsequently utilized to optimize a model capturing the underlying physical dynamics. 

\phuc{The integration of the WNCS with the DT model establishes a dynamic framework for real-time control and predictive analytics. This synergy enables the WNCS to leverage insights from the DT, facilitating precise control actions and proactive monitoring through closed-loop feedback mechanisms~\cite{almasan2022network}.} 
Given the interplay between communication and computation, a joint design strategy should preserve  the predictive power of the DT model, as well as execute accurate control signals, while simultaneously extending the operational lifespan of the network~\cite{ruah2023bayesian, 10092861}. 
In this regard, we introduce a DT architecture, as illustrated in Fig.~\ref{fig_system}(a), with the objective of addressing the ``what if'' query, including monitoring of what happened, what is happening with a certain degree of certainty, and predicting future outcomes upon execution of control signals. Monitored and constrained by specific requirements from the Operator, the DT model will optimize existing policies for more effective control and prediction.

As an illustration, consider an example of an Automated Guided Vehicle (AGV) operating within an indoor environment, such as a warehouse, where environmental conditions and the AGV's position and speed are continuously updated in a DT by acquiring sensory data. When the AGV moves through open spaces or is stationary, the system reduces the updating frequency to conserve wireless and computational resources. However, during high-speed movement, navigating tight corners, or encountering numerous obstacles, the system increases the frequency of updates to ensure precise and accurate control signals. This approach allows the DT model to effectively monitor the current state and predict future outcomes, optimizing resource usage while maintaining robust control and prediction capabilities. Within this framework, leveraging data and sensors accessible in the digital domain, we aim to harness partial sensory observations to forecast optimal policies for the system. Our proposed system architecture is tailored to the WNCS, wherein the DT deployed in the cloud is responsible for monitoring and estimating the system states within uncertainty requirements, geared towards recommending optimized control signals. This is done by adhering to stringent timing and reliability constraints. In order to minimize the communication overhead, the sensors are scheduled based on the quality of their observations as well as the communication constraints of the DT model.
\vspace{-5pt}
\subsection{Related Works}
In WNCSs, besides computational controlling methods \cite{kodagoda2002fuzzy, riazi2020energy}, there has been a substantial research on the data-driven method at the DT to perform various tasks~\cite{ruah2023bayesian, donmez2010probabilistic}. In related machine learning method such as active learning methodologies \cite{settles2011theories}, alongside their adaptations in Reinforcement Learning (RL) \cite{lopes2009active}, the principal objective revolves around reducing the training dataset size. This endeavor involves the agent strategically formulating informative inquiries during the training phase to optimize its performance during subsequent testing phases. Additionally, RL finds widespread application in policy optimization for control tasks, encompassing both model-based algorithms \cite{lopes2009active, dai2020deep} and model-free methods \cite{cronrath2019enhancing}. 
In these works, the RL agent assumes a passive role during testing, solely responding to queries presented to it, such as performing control when receiving noisy observation or image labeling tasks. 
In contrast to existing approaches, our framework maintains momentum during the testing phase by allowing the RL agent to process requests continuously. This decision is rooted in the necessity to gather data consistently, ensuring a steady stream of insights for refining and enhancing system performance. From this perspective, the greater the certainty required by the agent, the higher the cost incurred in terms of communication and computing resources. Consequently, we aim to balance minimizing the agent's inquiries and maintaining its ability to deliver precise control signals.

On the other hand, a critical challenge in WNCS lies in selecting sensors that effectively balance the trade-off between guaranteeing reliable state estimation and conserving sensor resources. In \cite{10092861}, the issue of scheduling IoT sensors is examined through the use of Value of Information (VoI) while taking into account the limitations of communication and reliability. \phuc{Fig.~1(a) highlights the necessity of implementing effective policies for selecting appropriate $\SA$s by considering their measurement inaccuracies, the demands of the DT model, and the costs associated with transmitting data over a wireless network.} The ultimate goal is to reduce the Mean Squared Error (MSE) of state estimation, which is often hindered by imprecise measurements. Additionally, authors in \cite{9768131} propose a solution for scheduling sensing agents through the utilization of VoI, ultimately leading to enhanced accuracy in a variety of summary statistics for state estimation.  These aforementioned works \cite{ruah2023bayesian,9768131, 10092861} and the references therein, however, do not consider the influence of control performance in the physical world and the strategy for selecting sensing agents based on the reliability of estimates and latency requirements. Moreover, the timeliness of knowledge within the system is assessed using the Age-of-Loop (AoL) metric, which serves to evaluate the overall age of a closed-loop WNCS \cite{de2023goal}. AoL can be considered as an extension of Age of Information (AoI), a well-established metric utilized for measuring information freshness in Internet of Things (IoT) applications \cite{champati2019performance}, which is primarily applicable to a single communication link, the DL or the UL \cite{yates2020age, gatsis2020latency }. In WNCSs such as the DT considered in this paper, the closed-loop communication creates mutual interdependence between DL and UL, which profoundly impacts the system performance and the optimal allocation of resources. 
\phuc{It is important to note that AoL directly influences control performance, as illustrated in Fig.~1(b). For timely state estimation at the DT model, the uplink sensing data transmission in the figure is subject to stringent latency constraints, i.e., ensuring that data is transmitted within a specific time frame. In addition, we also conduct a thorough assessment of the reliability of this transmission to guarantee its effectiveness and accuracy in diverse applications.}

\vspace{-5pt}
\subsection{Main Contributions}
Our main contributions are listed as follows: %\pp{PP: I think too many things are listed here, mixing up latency, AoL, VoI. Could this be restated in a more coherent manner, for example by using Figure 1 and explaining exactly what you want to achieve? In the example described above, in relation to Figure 1, you can introduce the roles of latency, AoL, and VoI.}
\begin{itemize}
    \item This work enhances  our preliminary DT architecture \cite{bui2024icc} for monitoring dynamic changes of the system parameters and controlling system dynamics under communication timing requirements in the form of reliability and Age of Information constraints. 
    This allows us to consider the overall impact of the communication system on the DT model and the physical world, and formulate the general problem of control and communication resource optimization.%, thereby optimizing the control and communication resources. 
    \item We introduce a RL framework to tackle the control problem, where the RL agent learns to perform the primary control tasks while dynamically adjusting observation certainty levels. By integrating these certainty requirements into the DT model, we develop an efficient scheduling policy that meets operational demands.
    \item We formulate a novel optimization problem to efficiently schedule $\SA$s for maintaining the confidence of DT's system estimate while minimizing the communication cost under latency and AoL requirements. The problem constraints encompass the necessary confidence levels expected from both the DT model and the RL agent to accurately compute an optimal control signal.
    \item \phuc{We propose a VoI-based algorithm that enables the scheduling of the most informative $\SA$s within polynomial time. We then derive a closed-form expression for the required physical resource block (PRB) for each selected $\SA$, under latency constraints and reliability thresholds.}
    \item Numerical simulations were conducted to evaluate the algorithm's performance, which confirms that it surpasses other benchmarks in both controlling  and power consumption while improving DT estimation error.
\end{itemize}

The rest of this paper is organized as follows: Section II presents the DT architecture, where the system model, communication diagram and problem formulation are detailed.  Subsequently, Section II presents a RL framework for optimizing control strategies while considering the uncertainty in estimating the system state. Building upon the requirements of latency and AoL within a limited communication budget, Section IV formulates and solves the problem of jointly minimizing bandwidth and scheduling $\SA$s with respect to uncertainty requirements from DT models and RL agent as in Section II. Utilizing the Extended Kalman Filter, we present an suboptimal heuristic scheduling strategy for $\SA$s and derive a closed-form for the optimal PRB allocation for active $\SA$s. Section V provides numerical results extensively, while main conclusions are finally given in Section VI.
\vspace{-5pt}
%%%%%%%%%%%%%%%%%%%%%%%%%%%%%%%%%%%%%%%%%%%%%%%%
\section{Digital Twin Architecture}
%%%%%%%%%%%%%%%%%%%%%%%%%%%%%%%%%%%%%%%%%%%%%%%%
\subsection{System Model}
We adopt a DT architecture, as illustrated in Fig.~\ref{fig_system}, to investigate our considered system. The physical world consists of a single primary agent ($\PA$) and a set of sensing agents ($\SA$s) denoted by $\Mcal=\{1,2,\dots, M\}$. These $\SA$s are responsible for observing the environment and establishing communication with the \emph{access point} (AP), which facilitates the construction of the DT model for the $\PA$. Each \emph{query interval} (QI) takes place at time instances $t \in \Tcal= \{1, 2, \dots, T\}$. To maintain synchronization and consistency, these $\mathtt{SA}$s periodically synchronize their DTs, encompassing information on locations and power budget status, with the Cloud platform, ensuring a high degree of reliability and managing the overhead of synchronization. Leveraging the current state of the DT model and the updated environmental information, the DT predicts the $\PA$'s state, devises an optimal policy for scheduling the $\SA$s, and generates optimal suggestions for subsequent actions. The feedback signal containing these insights is then conveyed to the AP, prompting it to execute appropriate actions in the physical world.

The primary agent ($\PA$) engages in interactions with the environment, operating within a $K$-dimensional process denoted as $\Kcal=\{1,2,\dots,K\}$. The state of this process at the $t$-th query interval (QI) is represented as $\mathbf{s}_t= [s_t^1, s_t^2, \dots, s_t^K]^\top$, and its evolution is described by the  relationship 
\begin{align}\label{dynamic_model}
    \mathbf{s}_t &= f(\mathbf{s}_{t-1}) + \bB a_{t-1} + \mathbf{u}_t, \quad \forall t\in\Tcal. 
\end{align}

% \bsa{COMMENT: n not defined} \phuc{I fixed it to $t$ and $\Tcal$}

Here, $f:\mathbb{R}^{K}\rightarrow\mathbb{R}^K$ denotes the state update function, $a_{t-1}$ is the control signal, and the matrix $\bB \in\mathbb{R}^K$ describes how the control impacts the dynamics. $\mathbf{u}_t\sim \mathcal{N}(\mathbf{0},\mathbf{C}_{\mathbf{u}})$ represents the process noise.
At QI $t$, each  $\SA$ $m \in \mathcal{M}$ receives a $D$-dimensional observation:
\begin{align}
    \mathbf{o}_{t,m} = g_m(\mathbf{s}_t)+ \mathbf{w}_{t,m},
\end{align}
corresponding to the $\PA$'s state and possesses $D\leq K$ dimensions. Herein, $g_m(\mathbf{s}_t)$ is the observation function of $\SA_m$. To facilitate simplicity in the analysis, the observation $\bo_{t,m}$ is assumed to be linearly dependent on the system state, which can be expressed as
\begin{align}
    \bo_{t,m}= \bH_m\bs_t+ \bw_{t,m}, \forall m\in\Mcal, 
\end{align}
where $\bH_m\in\mathbb{R}^{D\times K}$ is  the observation matrix, and $\bw_{t,m}\sim \Ncal(\mathbf{0}, \mathbf{C}_{\bw_m})$ stands for the measurement noise. In general, the covariance matrices $\mathbf{C}_{\bu}$ and $ \mathbf{C}_{\bw_m}$ are not diagonal.  We assume that $\bH_m$, $\mathbf{C}_{\bu}$ and $ \mathbf{C}_{\bw_m}$ are already known at the DT.

\vspace{-10pt}
\subsection{Communication System}
The communication model for the studied WCNS utilizes separate frequency bands for the uplink (UL) and downlink (DL) directions, resulting in independent medium access for UL and DL in the frequency domain. We assume  the communication link between AP and cloud is  perfect. Fig.~\ref{fig_system}(b) describe the communication diagram, where the parameter $T_\text{config}$ accounts for the time the DT model requires to update the active state of $\SA$s, after which it estimates the full system state, updates policy, computes optimal control signal, and schedules a maximum of $C$ $\SA$s via the AP using fusion algorithms and collected $\SA$s' data. Defining $\Qcal_t \subseteq \Mcal$ as the selected $\SA$ set at QI $t$, then the system must satisfy $|\Qcal_t| \leq C$.
Once a control command is generated, the controller promptly transmits it through a downlink channel to the physical world. To ensure close and reliable monitoring of the system, control signals are sent periodically after every QI, following a soft-sensor based control system approach \cite{fortuna2007soft}. On the physical world, the received command is stored in memory. After that, the application output for actuators control, such as motor drives, retrieves the most recent stored command values from memory and applies them to drive the system dynamics. Additionally, the scheduled $\SA$s read the system state and communicate the information back to the controller via the uplink channel. Both DL and UL transmissions are subject to potential latency during data delivery, which is influenced by two key factors: the channel quality and the total allocated PRBs for the transmission. 

The Fig.~\ref{fig_system}(b) illustrates the AoL behaviour of our considered representative time diagram, where the AoL starts at the DL, grows linearly over time  and drops at the time instances where the loop is closed ($t_{s1}, t_{s_2}, \dots$) to the corresponding timestamp in which the state feedback that a new control signal was generated, i.e., $\Delta L_t = t-t_{ci}, \forall i \in\{1,2,3,\dots\}$ \cite{de2023goal}. 
Given the presence of communication delay and packet losses (i.e., $\Delta L_t > 0$), our primary focus lies in investigating the asymptotic characteristics of the vehicle tracking error and the optimal control strategy. We define the contraint of AoL for keeping system up to date as
\begin{align}\label{AoL_require}
	\Delta L_{t,k} \leq \bar{\Delta}L_{k}, \forall k\in\Kcal,
\end{align}
\phuc{where $\bar{\Delta} L_t$ is the maximum tolerable AoL for feature $k\in\Kcal$ to keeping system stabilized, which is predetermined within the DT model.} 

Additionally, latency constraints for uplink communication are also taken into account, which ensures that observations will be transmitted successfully from $\SA$s to the AP. 
\phuc{The consideration of uplink latency is crucial because DT may be employed in applications beyond control that require timely state assessment. Moreover, $\SA$s, typically composed of IoT sensors, operate with low power and may have limited battery capacity. Therefore, accounting for uplink latency helps optimizing communication costs, including the bandwidth and/or  power  consumed by these $\SA$s.}
Specifically, let $\tau_{t,m}$ be the UL latency at $\SA_m$ and $\tau^\mathrm{max}$ the maximum tolerable UL latency for transmitting $\SA_m$'s information, the system fulfills the application reliability at a QI $t$ if
\begin{align}\label{aoi_requirement}
    \mathbb{P}[\tau _{t,m} >  \tau^\mathrm{max}] \leq \varepsilon, \forall m\in\Mcal, t\in\Tcal,
\end{align}
with $\varepsilon$ is a outage probability parameter depending on system characteristics \cite{8017572}.

\vspace{-5pt}
%%%%%%%%%%%%%%%%%%%%%%%%%%%%%%%%%%%%%%%%%%%%%%%%
\subsection{Problem Formulation}
The primary objectives of the DT model are to uphold a precise assessment of the state of a $\PA$ and offer the most advantageous sequence of actions to be executed in the physical realm based on its beliefs about states. A fundamental distinction within our system lies in its thorough consideration and evaluation of real-world environments, which often involve state observations that are characterized by noise or corresponded costs. Herein, the predicted estimator $\hat{\bs}_t$ of ${\bs}_t$ is modeled with 
\begin{align}
    p(\bs_t)\sim \mathcal{N}(\hat{\bs}_t, \bPsi_t), n \in\Ncal,
\end{align}
where the covariance matrix $\bPsi_t$ will be updated at QI $t$ according to the  Extended Kalman Filter (EKF). 
The MSE of the estimator is then %$\text{MSE}_{} = \mathbb{E}\big[||\bs_t-\hat{\bs}_t||^2_2\big], n \in\Ncal$. 
\begin{align}\label{MSE_EKF}
    \text{MSE}_{} = \mathbb{E}\big[||\bs_t-\hat{\bs}_t||^2_2\big], n \in\Ncal. 
\end{align}

\begin{remark}\label{certainty}
    We define the maximum acceptable standard deviation for feature $k\in\Kcal$ as $\xi_k$. This corresponds to the following condition:
    \begin{equation}\label{qos_condition}
        [\bPsi_t]_k \leq {\xi}_k^2, \forall k \in\Kcal,
    \end{equation}
    where $[\bPsi_t]_{k}$ is the $k$-th element of the diagonal of $\bPsi_t$.% \forest{define this as well}.
\end{remark}
Defining  $V^\pi(\bs_0)$ as value function of controlling $\PA$ in \eqref{dynamic_model} under control policy $\pi$, 
we are interested in jointly minimizing the sum PRB consumption and delivering optimal control signals.
\phuc{Before determining the allocation of PRBs to scheduled $\SA$s, it is essential to accurately compute the required bandwidth. This ensures optimal utilization of resources, respecting the fact that each active  $\SA$ will utilize its entire power budget for packet transmission. By advancing this, we can make informed decisions about bandwidth allocation, ensuring conditions on latency and outage probability as stated in \eqref{qos_condition}. Denoting $W_{t,m}$ as the bandwidth scheduled for $\SA_m$ at QI $t$, we introduce}
\begin{IEEEeqnarray}{ll}\label{glob_objective}
	h(\{W_{t,m}\}, \{\ba_t\}) = [V^\pi(\hat{\bs}_0), -\sum_{t=0}^{\infty}\sum_{m=1}^{M}W_{t,m}]^\top,
\end{IEEEeqnarray}
which is categorized as a multi-objective function, where the two performance metrics are optimized in a single framework. Motivated by the use of \eqref{glob_objective}, we formulate the  optimization problem  as a joint design of control and scheduling $\SA$s to minimize the bandwidth consumption while maintaining the confidence of DT's system estimate under timing requirements. At each QI, the problem is formulated as
\begin{subequations} \label{glob_problem}
    \begin{alignat}{2}
	\text{P1}: \underset{\{W_{t,m}\}, \{\ba_t\}}  { \mathrm{maximize}} \ & h\big(\{W_{t,m}\}, \{\ba_t\}\big) \label{glob_problema}\\
	\mathrm{s.t.} \quad  & \mathbf{s}_t = f(\mathbf{s}_{t-1}) +  \bB \ba_{t-1} + \mathbf{u}_t, \label{glob_problemb}\\
   & \mathbb{P}[\tau _{t,m} >  \tau^\mathrm{max}] \leq \varepsilon, \forall m\in\Mcal, t\in\Tcal,\label{glob_problemc}\\
   & [\bPsi_t]_k \leq {\xi}_k^2, \forall k \in\Kcal, t\in\Tcal, \label{glob_problemd}\\
   & \Delta L_{t,k} \leq \bar{\Delta}L_{k}, \forall t\in\Tcal,  k\in\Kcal,\label{glob_probleme}\\
    &|\Qcal_t| \leq C, \forall t\in \Tcal.\label{glob_problemf}
    \end{alignat}
\end{subequations}
In the problem \eqref{glob_problem}, the constraint \eqref{glob_problemb} describes the system dynamics, while \eqref{glob_problemc} indicates the timing constraint of uplink communication. The certainty requirement of estimation process is shown in \eqref{glob_problemd}. The constraint \eqref{glob_probleme} guaranteeing the maximum tolerable AoL for feature $k\in\Kcal$ does not exceed the threshold $\bar{\Delta}L_k$ to keeping the system stabilised. Finally, \eqref{glob_problemf} holds the condition about uplink connections. 
It is noted that we consider a scenario where the belief vector can be enhanced through estimation techniques facilitated by DT prior to being utilized by RL to suggest the optimal action as an output. \phuc{This necessitates devising an optimal strategy for allocating PRBs to strike a balance between satisfying the objectives \eqref{glob_problema} and the constraints imposed for AoL \eqref{glob_probleme} and latency \eqref{glob_problemc}.}
By employing more $\SA$s, the accuracy of estimated state can be enhanced, enabling the agent to make more precise decisions. Moreover, if the estimator can consistently estimate the system's state with reliability, the need to gather supplementary observations from $\SA$s becomes unnecessary, resulting in direct savings in terms of communication and information processing.

In the following sections,  we propose the AoL-REVERB (AoL-based REinforcement learning and Variational Extended Kalman filter with Robust Belief) framework including two-step approach to address the problem \eqref{glob_problem}: $(i)$, we employ an uncertainty control RL algorithm to devise control actions aimed at aplying to the physical world and effectively managing the observation errors received from $\SA$s; and $(ii)$, the $\SA$ scheduling algorithm based on VoI and AoL considerations and optimal PRB control algorithm are utilized to identify the most significant  $\SA$s for observing its sensing signals, guided by the requirements from the RL model and DT. To further enhance the accuracy of estimated states and forecast the forthcoming system state,  the EKF technique is revised and integrated. \phuc{The main concept is that the $\SA$s contributing the most significant observations to satisfied AoL constraints \eqref{AoL_require} and reduce the MSE in \eqref{MSE_EKF} are prioritized, especially when the accuracy of the $\PA$'s future state estimates based on the EKF in \eqref{qos_condition} is not yet satisfied. Then, the optimal allocated PRBs for those scheduled $\SA$s, ensuring reliable uplinks as stated in \eqref{aoi_requirement}, are provided in a closed-form manner.}
It is remarkable that the successful application of the EKF technique necessitates awareness of the system's transistion, which might involve resorting to mathematical methods or optimization techniques to determine optimal/suboptimal actions.

\vspace{-5pt}
%%%%%%%%%%%%%%%%%%%%%%%%%%%%%%%%%%%%%%%%%%%%%%%%
\section{Uncertainty Control POMDP}
%%%%%%%%%%%%%%%%%%%%%%%%%%%%%%%%%%%%%%%%%%%%%%%%
\vspace{-0pt}
\phuc{For finding optimal control policy $\pi$ outlined in \eqref{glob_problem}, alongside methods utilizing mathematical models or computational control signals, RL presents a promising approach for addressing control issues, especially when computing optimal control becomes challenging in conditions of system noise and complex environments \cite{ruah2023bayesian}. By executing RL algorithms, an autonomous agent can acquire optimal control strategies through interaction with its environment.}
While RL algorithms are commonly formulated in terms of Markov Decision Processes (MDPs), we note that in typical real-life applications, the states are often unobservable directly. In other words, the observations provided only offer partial and potentially inaccurate information \cite{murphy2000survey}. 
Therefore, we consider the problem \eqref{glob_problem} as Partially Observable Markov Decision Process (POMDP) that expands upon the MDP by incorporating the sets of observations and observation probabilities to actual states.

In particular, a POMDP is presented by a 7-tuple $ \langle \Scal, \Acal, \Ocal, \mathtt{P}, \mathtt{O}, r ,\gamma \rangle$. In particular, $ \Scal $ is the finite set of possible states, $ \Acal $ is a  set of control primitives and $ \Ocal $ denotes a  set of possible observations. At a time instant $ t $, the agent makes an action $\ba_t$ to move from state $ \bs_t $ to $ \bs_{t+1} $ with the transition probability  $ \Ptt = \mathbb{P}[\bs_{t+1}|\bs_{t},\ba_t]$. An observation $ \bo_{t+1} $ received from $\SA$s tracking system's state occurred with a probability $ \Ott = \mathbb{P}[\bo_{t+1}|\bs_{t+1},\ba_t] $. Also, upon transition the agent receives a numerical reward $ r(\bs_{t}, \ba_t, \bs_{t+1}) $ verifying	$ r(\bs_{t},\ba_t,\bo_{t+1}) \leq r^\mathrm{max} $. 
An agent does not know exactly its state at QI $t$ and it maintains an estimate-vector $\hat{\bs}_t$ describing the probability of being in a particular state $ \bs_t \in\Xcal $. 
We define $ \pi $ as a policy of the agent that specifies an action $ \ba_t $ based on its policy  $\pi(\hat{\bs}, \ba)$. 
Start from initial belief $\hat{\bs}_0$, the expected future discounted reward for policy $\pi(\hat{\bs},a)$ is given as
\begin{equation}\label{}
    V^\pi(\hat{\bs}_0) = \mathbb{E}\left[\sum_{t=0}^{\infty}\gamma_t r(\bs_{t},\ba_t,\bs_{t+1}) |\hat{\bs}_0,\pi\right],
\end{equation}
where $0< \gamma_t<1 $ is the discount factor.  In the system, the goal for the agent is to make action sequence $ \{\ba_t\} $ that maximizes the long-term reward. In other word, the agent try find the optimal policy $\pi^*$ that satisfies
\begin{equation}\label{RL_problem}
    \pi^* = \underset{\pi}{\arg\max }\ V^\pi(\hat{\bs}_0).
\end{equation}

\begin{center}
	\begin{figure*}[ht]
		\centering
		\includegraphics[trim=-2cm 0cm 0.cm 0cm, clip=true, width=1.0\textwidth]{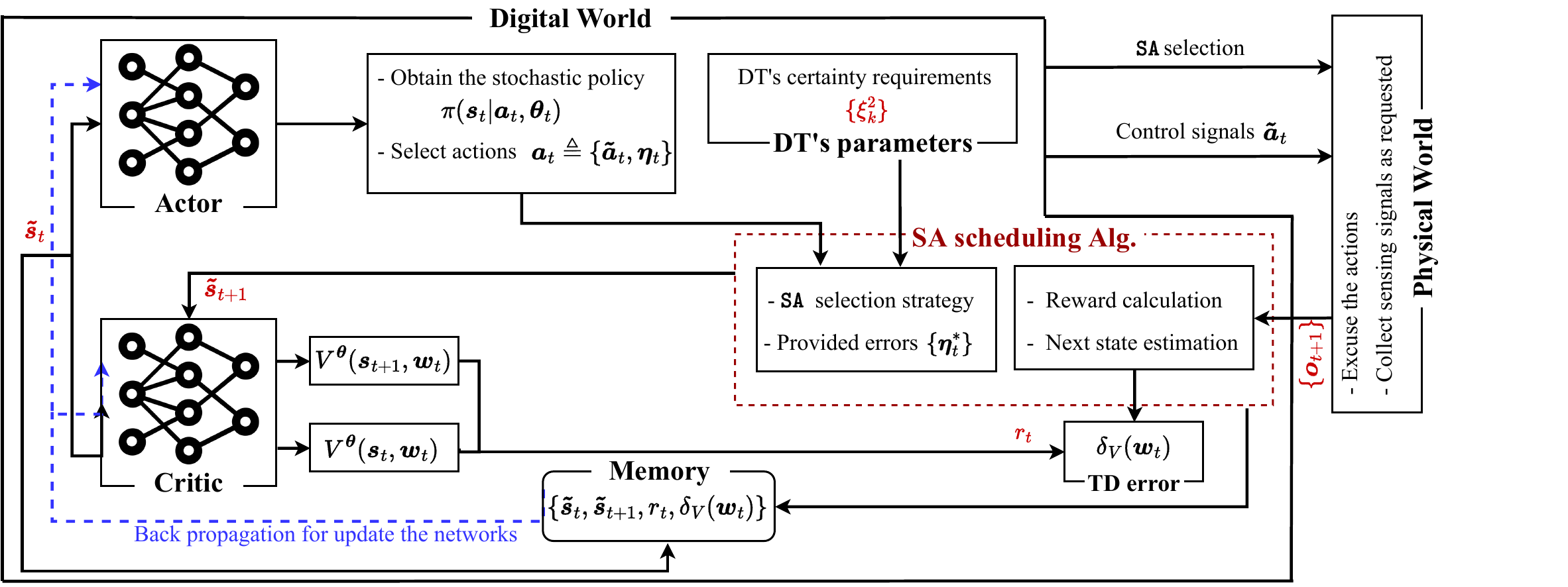}
		\caption{The  actor-critic framework of ACEC algorithm.}
		\label{fig_technique}
		\vspace*{-0.55cm}
	\end{figure*}
\end{center}

At QI $t$, the estimated state vector is given as $\hat{\bs}_t  = [\hat{\bs}_t^1, \hat{\bs}_t^2, \dots, \hat{\bs}_t^K]^\top  \in \mathbb{R}^K$. The estimated stated can be governed by 
\begin{align}
    \hat{\bs}_t \sim p(\hat{\bs}_t|\bs_t; \boldsymbol{\eta}_t), 
\end{align}
where $ p(\hat{\bs}_t|\bs_t; \boldsymbol{\eta}_t)$ is the  conditional probability distribution function (pdf) of $\hat{\bs}_t$ given $\bs_{t}$. 
Specially, $ \boldsymbol{\eta}_t =[ {\eta}_{t,1},  {\eta}_{t,2}, \dots, {\eta}_{t,K}]^\top \in \mathbb{R}^K$ indicates the average accuracy vector, whose each element $\eta_{t,k}$ indicates the accuracy of $k$-th process of estimated state $\hat{\bs}_t$.
In this work, we define 
\begin{align}\label{eta_compute}
    {\eta}_{t,k} = \frac{1}{[\bPsi_t]_{k}} , \forall k\in \Kcal.
\end{align}
At the results, we could parametrize $ p(\bs_t) \sim \mathcal{N}(\hat{\bs}_t, \diag{(\boldsymbol{\boldsymbol{\varpi}}_t)} = \bPsi_t)$, with $\boldsymbol{\varpi}_t = [\frac{1}{\eta}_{t,1}, \frac{1}{\eta}_{t,2}, \dots, \frac{1}{\eta}_{t,K}]$. 
It is evident that as  ${\eta}_{t,k} $ increases, the confidence level of $\hat{\bs}_{t,k}$ also increases, facilitating the RL agent in making accurate decisions. Nevertheless, achieving high reliability of $\hat{\bs}_{t,k}$ necessitates the usage of corresponding $\SA$s with low error of measurements or more than one observation, consequently amplifying both communication and processing expenses.
Given $\bs_{t}$ and $\boldsymbol{\eta}_t$, the estimated state could be assumed exhibit statistical independence, meaning that we can express 
\begin{align}
    p(\hat{\bs}_t|\bs_t; \boldsymbol{\eta}_t) = \prod_{k\in \Kcal} p(\hat{\bs}_{t,k}|\bs_t; \boldsymbol{\eta}_t)
\end{align}
in terms of their factorization. Here, $p(\hat{\bs}_{t,k}|\bs_t; \boldsymbol{\eta}_t)$ represents the conditional pdf of $\hat{\bs}_{t,k}$ given $\bs_t$ and $ \boldsymbol{\eta}_t$. 

\vspace{-10pt}
\subsection{Actor-critic-based DRL Algorithm }
For training the policy $\pi(\boldsymbol{\hat{s}},\boldsymbol{a})$, this study employs Proximal Policy Optimization (PPO) \cite{schulman2017proximal}, a well-established reinforcement learning (RL) algorithm, to achieve its objectives. Within the realm of RL, the actor-critic structure is widely adopted for agents that involves dividing the model into two distinct components, thus harnessing the strengths of both value-based and policy-based methods \cite{NIPS1999_6449f44a}. Specifically, the actor is primarily responsible for estimating the policy, which dictates the agent's actions in a given state, while the critic is dedicated to estimating the value function, which predicts the expected future reward for a particular state or state-action pair. Moreover, the actor's policy undergoes refinement based on the feedback provided by the critic. In the context of the stochastic policy $\pi(\ba_t|\hat{\bs}_t)$, the actor takes the responsibility of selecting actions, and subsequently, the critic component is utilized to evaluate these decisions through a Q-value, denoted by $Q^{\pi}(\hat{\bs}_t,\ba_t)$, which can be mathematically expressed as 
\begin{align}
    Q^{\pi}(\hat{\bs}_t,\ba_t)=\mathbb{E}_{\ba_t\sim\pi(\ba_t\hat{\bs}_t)}[R_t|\hat{\bs}_t, \ba_t],
\end{align}
where $\mathbb{E}_{\boldsymbol{a}_t\sim\pi(\boldsymbol{a}_t|\hat{\bs}_t)}[\centerdot|\centerdot]$ represents a conditional expectation under  $\pi(\boldsymbol{a}_t|\hat{\bs}_t)$, and $R_t$ corresponds to  the cumulative discounted reward with a discount factor $\gamma$,  expressed as:
\begin{align}
    R_t = \sum_{t'=t}^{\infty}\gamma^{t'-t}r_{t'},~~\gamma\in [0,1].
	\end{align}
In practical scenarios, obtaining explicit expressions for $\pi(\ba_t|\hat{\bs}_t)$ and $Q^{\pi}(\hat{\bs}_t, \ba_t)$ is complex and computationally expensive. Therefore, Deep Neural Networks (DNNs) are employed as parameterized approximators to furnish estimations for $\pi(\ba_t|\hat{\bs}_t)$ and $Q^{\pi}(\hat{\bs}_t, \ba_t)$.
Let $\boldsymbol{\theta}_t$ and $\boldsymbol{\omega}_t$ be the parameter vectors associated with the actor and critic, and denote $\pi(\boldsymbol{a}_t|\hat{\bs}_t;\boldsymbol{\theta}_t)$ and $Q^{\boldsymbol{\theta}}(\hat{\bs}_t,\boldsymbol{a}_t;\boldsymbol{\omega}_t)$ as the respective parameterized functions. To simplify the notation, we express $Q^{\boldsymbol{\theta}}(\hat{\bs}_t,\boldsymbol{a}_t;\boldsymbol{\omega}_t)$ as $\mathbb{E}_{\boldsymbol{a}_t\sim\pi(\boldsymbol{a}_t|\hat{\bs}_t;\boldsymbol{\theta}_t)}[R_t|\hat{\bs}_t, \boldsymbol{a}_t]$.
The objective is to minimize the loss function of the actor, denoted as $-J(\boldsymbol{\theta}_t)$, which is expressed as $-J(\boldsymbol{\theta}_t) = -\mathbb{E}[Q^{\boldsymbol{\theta}}(\hat{\bs}_t,\boldsymbol{a}_t;\boldsymbol{\omega}_t)]$. 
By leveraging the fundamental outcomes of the policy gradient theorem \cite{sutton2018reinforcement}, the gradient of $J(\boldsymbol{\theta}_t)$ can be computed as
\begin{align}
    \label{eq:policy_gradient}
    \nabla_{\boldsymbol{\theta}}J(\boldsymbol{\theta}_t) = \mathbb{E}[\nabla_{\boldsymbol{\theta}}\log\pi(\boldsymbol{a}_t|\hat{\bs}_t;\boldsymbol{\theta}_t)Q^{\boldsymbol{\theta}}(\hat{\bs}_t,\boldsymbol{a}_t;\boldsymbol{\omega}_t)].
\end{align}
The update rule of $\boldsymbol{\theta}_t$ can be derived based on gradient descent:
\begin{align}
    \label{update_actor}
    \boldsymbol{\theta}_{t+1} =  \boldsymbol{\theta}_{t} - \alpha_a\cdot(\nabla_{\boldsymbol{\theta}} J(\boldsymbol{\theta}_{t})),
\end{align}
where $\alpha_a$ is the learning rate of the actor. 

On the critic's side, the parameter vector $\boldsymbol{\omega}_t$ undergoes updates through temporal-difference (TD) learning \cite{sutton2018reinforcement}. In the context of TD learning, the loss function for the critic, denoted as $C{_Q}(\boldsymbol{\omega}_t)$, is formally defined as the expectation of the squared TD error, $\delta{_Q}(\boldsymbol{\omega}_t)$, which is expressed as $\mathbb{E}[(\delta{_Q}(\boldsymbol{\omega}_t))^2]$.
The TD error $\delta_{_Q}(\boldsymbol{\omega}_t)$ pertains to the discrepancy between the TD target and the estimated Q-value, and it is  expressed as
\begin{align}
    \label{TD_Q}
    \delta_{_Q}(\boldsymbol{\omega}_t) = r_t+\gamma Q^{\boldsymbol{\theta}}(\hat{\bs}_{t+1},\boldsymbol{a}_{t+1};\boldsymbol{\omega}_t)-Q^{\boldsymbol{\theta}}(\hat{\bs}_t,\boldsymbol{a}_t;\boldsymbol{\omega}_t),
\end{align}
where $r_t+\gamma Q^{\boldsymbol{\theta}}(\hat{\bs}_{t+1},\boldsymbol{a}_{t+1};\boldsymbol{\omega}_t)$ represents the TD target.
The primary objective of the critic is to minimize the loss function $C_{_Q}(\boldsymbol{\omega}_t)$, and the update rule for $\boldsymbol{\omega}_t$ can be obtained through gradient descent as 
\begin{align}
    \label{update_critic}
    \boldsymbol{\omega}_{t+1} = \boldsymbol{\omega}_t - \alpha_c\nabla_{\boldsymbol{\omega}} C_{_Q}(\boldsymbol{\omega}_t),
\end{align}
where $\alpha_c$ represents the learning rate employed for the critic.

It is worth noting that the approximation of $Q^\pi(\hat{\bs}_t,\boldsymbol{a}_t)$ introduces a significant variance in the gradient $\nabla_{\boldsymbol{\theta}}J(\boldsymbol{\theta}_t)$, leading to suboptimal convergence \cite{schulman2015high}. To address this issue, a V-value is introduced by
\begin{align}
    V^\pi(\hat{\bs}_t) =\mathbb{E}_{\boldsymbol{a}_t\sim\pi(\boldsymbol{a}_t|\hat{\bs}_t)}[R_t|\hat{\bs}_t].
\end{align}
By approximating $V^\pi(\hat{\bs}_t)$, it is possible to reduce the variance.
We employ the parameterized V-value $V^{\boldsymbol{\theta}}(\hat{\bs}_t; \boldsymbol{\omega}_t)$, then the TD error and the loss function for the critic can be respectively expressed as 
\begin{align}
    \label{eq:TD_V}
    \delta_{_V}(\boldsymbol{\omega}_t) = r_t+\gamma V^{\boldsymbol{\theta}}(\hat{\bs}_{t+1};\boldsymbol{\omega}_t)-V^{\boldsymbol{\theta}}(\hat{\bs}_t;\boldsymbol{\omega}_t),
\end{align}
and 
\begin{align}
    \label{eq:loss_TD_V}
    C_{_V}(\boldsymbol{\omega}_t) = \mathbb{E}[(\delta_{_V}(\boldsymbol{\omega}_t))^2].
\end{align}
Moreover, it is important to note that $\delta_{_V}(\boldsymbol{\omega}_t)$ provides an unbiased estimation of the Q-value \cite{schulman2015high}. Consequently, we can rewrite $\nabla_{\boldsymbol{\theta}} J(\boldsymbol{\theta}_t)$ from Eq. (\ref{eq:policy_gradient}) by
\begin{align}
    \nabla_{\boldsymbol{\theta}} J(\boldsymbol{\theta}_t) =& \mathbb{E}\left[\nabla_{\boldsymbol{\theta}}\log(\pi(\boldsymbol{a}_t|\hat{\bs}_t;\boldsymbol{\theta}_t))Q^{\pi}(\hat{\bs}_t,\boldsymbol{a}_t)\right]\notag\\
    =&\mathbb{E}\left[\nabla_{\boldsymbol{\theta}}\log(\pi(\boldsymbol{a}_t|\hat{\bs}_t;\boldsymbol{\theta}_t))\delta_{_V}(\boldsymbol{\omega}_t)\right].
\end{align}	
To address the joint design problem with the objective of optimizing actions while minimizing communication energy, we implement the Actor-Critic Deep Reinforcement Learning (DRL) approach within the DT cloud environment, where  the  actions and rewards need to be redefined.

\vspace{-5pt}
\subsection{Action Space Reformulation}
We focus on the issue of optimizing the accuracy of the estimated state by the RL agent, enabling it to select $\eta_{t,k}$ on a continuous scale. The underlying motivation for the proposed framework is to unveil the inherent characteristics of the observation space in terms of the informational value the observations offer for the given task. Furthermore, this framework can serve as a RL-based solution for addressing decision-making problems in real-world scenarios, particularly when observations typically entail associated costs such as equipment expenses related to measurement accuracy and the communication overhead.
	
The aforementioned discussions culminate in the formulation of the action vector structure within the RL agent implemented in DT, represented as
\begin{align}\label{action_space}
    \ba_t  = [{a}_{t,1},\dots,{a}_{t,Z},\eta_{t,1},\dots,\eta_{t,K}] \in\mathbb{R}^{ZK},
\end{align}
where $\mathcal{Z}$ is the action space with $(|\mathcal{Z}|= Z)$. In \eqref{action_space}, $\{{a}_{t,z}\}_{z\in\mathcal{Z}}$ correspond to the control signals that exert an influence on the physical environment, enabling the agent to advance towards its objective. Additionally,  $\eta_{t,k}\in[0,\infty]$ denotes the accuracy selection pertaining to the estimated state.
	
\vspace{-5pt}
\subsection{Reward Function Reformulation}
It is imperative for the RL agent to not only navigate towards the primary objective defined for the problem \eqref{RL_problem} but also acquire the ability to regulate the acceptable level of accuracy  $\{\eta_{t,k}\}_{k\in\Kcal}$. Consequently, the goal-based reward $r_t$ is transformed into an uncertainty-based reward as $\tilde{r}_t = f(r_t, \boldsymbol{\eta}_t),$ 
wherein $f(\cdot)$ is a monotonically non-decreasing function of $r_t$ and $\boldsymbol{\eta}_t$. 
In scenarios where a direct cost function, denoted as $c_k(\cdot)$, exhibits an upward trend with the accuracy of the observation $o_{t,k}$, a suitable additive formulation can be employed. Specifically, the modified reward, $\tilde{r}_t$, can be expressed by \cite{koseoglu2020learning}
\begin{align}\label{reward_shaping}
    \tilde{r}_t = r_t + \kappa  \sum_{k=1}^K c_k(\eta_{t,k}).
\end{align}
Here, $c_k(\eta_{t,k})$ represents a non-increasing function of $\eta_{t,k}$, and $\kappa \geq 0$ serves as a weighting parameter. Therefore, the objective of the agent is two-fold: to maximize the original reward while simultaneously minimizing the cost associated with the observations. This trade-off can be intuitived through the following  example:
\begin{example}\label{example1}
    In the conventional RL problem known as MountainCar \cite{Moore90efficientmemory-based}, a car strives to reach a designated position where a flag is located on top of the right hill, with a velocity within $[-\infty, +\infty]$, as depicted in Fig.~\ref{fig_trajectory}(a). When the car is situated far from the flag's position, a force can be applied to propel the car forward without requiring precise knowledge of its exact position and velocity. In this scenario, gathering numerous observations to accurately estimate the state of the system becomes redundant and inefficient in terms of resource utilization. However, as the car approaches the flag and/or enters a critical position, such as being on a slope, precise information regarding the coordinates and current position of the car becomes necessary to apply a suitable and subtle force.
	\end{example}
	
Our proposed algorithm is summarized in \textbf{Algorithm~\ref{AC_Alg}}, where the flow is illustrated in Fig.~\ref{fig_technique}.

\begin{algorithm}[t]
	\begin{algorithmic}[1]{\fontsize{9pt}{9pt}\selectfont
			\protect\caption{ACEC (Actor-Critic Error Controlling) algorithm }% \eqref{probGlobal}}
		\label{AC_Alg}
		\global\long\def\algorithmicrequire{\textbf{Input:}}
		\REQUIRE Current estimated state $\hat{\bs}_t$
		\global\long\def\algorithmicrequire{\textbf{Output:}}
		\REQUIRE Current action $\ba_t$ and required error $\boldsymbol{\eta}_t$
		\STATE Randomly initialize critic network and actor network with weights $\boldsymbol{\omega}_1$ and $\theta_1$, respectively
		\FOR{each learning round} 
		\STATE Receive initial estimated state $\hat{\bs}_0$ from DT
		\FOR{$t = 1:T$}
		\STATE Approximate a distribution $\pi(\ba|\bs_{t},\theta_t) $ by actor
		\STATE Sample action $\ba_t, \boldsymbol{\eta}_t \sim \pi(\ba|\bs_{t},\theta_t) $ and execute to the physical world
		\STATE Send $\boldsymbol{\eta}_t$ to the DT
		\STATE Observe reward $r_t$ as in \eqref{reward_shaping}
		\STATE Collect next estimated state $\hat{\bs}_{t+1}$  from DT
		\STATE Pass $r_t$ and $\hat{\bs}_{t+1}$ to the critic   
		\STATE Approximate $Q(\ba_t, \hat{\bs}_{t}|\boldsymbol{\omega}_t), Q(\ba_{t+1}, \hat{\bs}_{t+1}|\boldsymbol{\omega}_{t+1})$	by critic 
		\STATE  Calculate TD error $\delta_{_Q}(\boldsymbol{\omega}_t)$ by \eqref{TD_Q}
		\STATE Update $\theta_t$ and $w_t$ as in \eqref{update_actor} and \eqref{update_critic}, respectively	
		\ENDFOR
		\ENDFOR
	}
\end{algorithmic}
\end{algorithm}

%%%%%%%%%%%%%%%%%%%%%%%%%%%%%%%%%%%%%%%%%%%%%%%%
\section{$\SA$s Scheduling and PRB Allocation}
%%%%%%%%%%%%%%%%%%%%%%%%%%%%%%%%%%%%%%%%%%%%%%%%

In this section, our focus is on the DT domain, where the task involves the scheduling of $\SA$s based on three pivotal factors: $(i)$ the acceptable level of accuracy for the estimated state, as determined by the RL agent, plays a crucial role;  $(ii)$ the requirement pertaining  to the accuracy  of the DT model as in \eqref{qos_condition}, ensuring that it sustains the tracking of the physical system within a dependable threshold ; and $(iii)$ the communication resources, determined by the system capacity, latency, and reliability, and AoL requirements. 

\vspace{-5pt}
\subsection{Sensing Agent Scheduling Problem }

Subsequently, we formulate a combined $\SA$s scheduling and bandwidth control problem, where our goal is to determine the $\SA$s that should engage in transmission for minimizing power consumption, all the while satisfying the prescribed reliability $\{\xi_k^2\}_{k\in\Kcal}$ outlined in the  constraint  \eqref{qos_condition} and the required accuracy $\boldsymbol{\eta}_t$ at RL agent. We propose using arbitrary variables $\bar{\xi}_{t,k}^2$  standing for desired DT's error level of system's state at QI $t$.  Given the necessary reliability stipulated for the DT as presented in \eqref{qos_condition} and the requisite level of accuracy $\boldsymbol{\eta}_t$ to uphold the precision of the RL model, the DT should meet the error constraints at QI $t$ as
\begin{equation}\label{glob_qos}
	[\bPsi_t]_k \leq \bar{\xi}_{t,k}^2 \triangleq \min\Big\{\xi_k^2, \frac{1}{\eta_{t,k}}\Big\}, \forall t\in\Tcal, k\in\Kcal.
\end{equation}
Initially, we establish the reachable $\SA$ set $\Pcal_t$ at QI $t$ as $\Pcal_t = \Mcal$. 
The problem is thus formulated with the following specifications, utilizing the bandwidth allocation vector $\{W^\text{tx}_{t,m}\}_{m\in\Mcal}$ as the optimization variable: 
\begin{subequations} \label{scheduling_problem}
    \begin{alignat}{2}
        \{W_{t,m}^{\mathrm{tx}*}\} &= \  \underset{\{W_{t,m}^{\mathrm{tx}}\}}  { \mathrm{argmin}} \ & &(1-\alpha) \sum_{k\in\Kcal}\max\left\{\frac{[\bPsi(n)]_{k }}{\bar{\xi}^2_{k}}-1,0\right\} \nonumber\\
        &&&+\alpha \sum_{\mathclap{m\in\Pcal(n)}}W^\mathrm{tx}_m(n)\label{scheduling_problema}\\
        &\mathrm{subject\ to }\  && |\Qcal_t| \leq C, \forall t\in \Tcal, \label{scheduling_problemb}\\
        &&& \mathbb{P}[\tau_{t,m} > \tau^\mathrm{max}] \leq \varepsilon, \forall t\in \Tcal,  m\in\Mcal,  \label{scheduling_problemc}\\
		&&&\Delta L_{t,k} \leq \bar{\Delta}L_{k}, \forall t\in\Tcal,  k\in\Kcal, \label{scheduling_problemd}
    \end{alignat}
\end{subequations}
wherein the non-negative parameter $\alpha\in[0,1]$ represents the relative weight to accuracy and energy efficiency within the underlying objective function. It is observed that objective \eqref{scheduling_problema} represents a relaxation of constraint \eqref{glob_qos} due to its dependence on prevailing conditions, i.e., in situations where the error surpasses a certain threshold, even the complete assimilation of sensor data fails to guarantee the desired level of reliability $\bar{\xi}^2_{t,k}$. Under the condition that this requirement is fulfilled, the strict imposition of constraint \eqref{glob_qos} results in the infeasibility of problem \eqref{scheduling_problem} since it leads to an empty feasible set.
It is easy to see that obtaining observations from additional sources $\SA$s leads to an enhancement in estimation accuracy; however, this improvement comes at the cost of compromising energy efficiency. For those particular $\SA$s that exhibit considerable errors in their measurements or possess features that do not significantly contribute to meeting the confidence requirements of the $\PA$ (i.e., those with a low VoI), measuring and transmitting observations leads to an unnecessary expenditure of energy. The constraints given by \eqref{scheduling_problemb} and \eqref{scheduling_problemc} are required in ensuring the reliable execution of transmissions. These constraints impose limitations such that not more than $C$ $\SA$s can engage in transmission during any given QI. Additionally, each individual $\SA$ possesses the scheduling to either transmit data with power $\bar{p}^\text{tx}$ or enter a standby mode. It is important to underscore that the optimization problem presented as \eqref{scheduling_problem} is inherently non-convex due to the non-convex nature of the objective function \eqref{scheduling_problema}, as well as the the constraint \eqref{scheduling_problemb} and binary selection of \eqref{scheduling_problemc}. Furthermore, the node selection aspect renders the problem analogous to the classic NP-hard knapsack problem. Consequently, to derive an efficient suboptimal solution, a heuristic algorithm based on EKF is employed.

\begin{algorithm}[t]
	\begin{algorithmic}[1]{\fontsize{9pt}{9pt}\selectfont
			\protect\caption{$\SA$ scheduling algorithm for problem  \eqref{scheduling_problem}} % \eqref{probGlobal}}
		\label{scheduling_alg}
		\global\long\def\algorithmicrequire{\textbf{Input:}}
		\REQUIRE $\bb_{0,\bo_0}, \bmu_{\bu_0}, C_{\bu_0}$
		Available uplink slots $C$,  The state and requirement certainty $\big(\bs, \{\bar{\xi}_{k}^2\}\big)$
		\global\long\def\algorithmicrequire{\textbf{Output:}}
		\REQUIRE The scheduled user set $\{\Qcal^*_t\}$; their belief $\{\hat{\bs}^*_t, \bPsi^*_t\}$, and their associated PRBs.
		\STATE Initialize $\mathcal{Q}_t = \emptyset$
		\STATE Compute the prior errors $\bPsi^{\mathtt{pr}}_t$ as in \eqref{prior_error} and solving the constraint \eqref{scheduling_problemd} with \eqref{AoL_update}
		\IF {$[\bPsi^{\mathtt{pr}}_t]_{k} \leq \xi_{k}^2, \forall k$}
		\STATE Compute ${\hat{\bs}^{\mathtt{pr}}_t} $ as in \eqref{mu_prior}
		\STATE Update ${\hat{\bs}_t}  = {\hat{\bs}^{\mathtt{pr}}_t} $ and $\bPsi_{\hat{\bs}_t} = \bPsi_{\hat{\bs}_t}^\mathtt{pr}$
		\ELSE
		\STATE Set $i=1$ 
		\WHILE{conditions \eqref{checking3_conditions} hold}
		\STATE Update $\Qcal_t$ and $\Pcal_t$ as in \eqref{update_set}
		\STATE Update the $\bK_t$, $\bH_t$ and $\bC_{\bw_t}$ as in \eqref{H_compute}, and \eqref{Cw_compute}
		\STATE Compute $	\bPsi^{\mathtt{pos}}_{\hat{\bs}_t}$ as in \eqref{pos_variance}
		\STATE  Set $i=i+1$
		\ENDWHILE
		\STATE Update $\Qcal^*_t = \Qcal_t$
		%		\STATE Compute $\bH_t$ and $\bC_{\bw_t}_t$ as in \eqref{H_compute} and \eqref{Cw_compute}
		\STATE Compute ${\hat{\bs}_t}  = {\hat{\bs}^{\mathtt{pos}}_t} $ and $\bPsi_{\hat{\bs}_t} = \bPsi_{\hat{\bs}_t}^\mathtt{pos}$ as in\eqref{pos_update},   \eqref{pos_variance}
		\STATE Update $\boldsymbol{\eta}^*_t$ as in \eqref{eta_compute}
		\STATE Update the bandwidth transmission as in \eqref{optimal_bandwidth}, then compute the PRBs as in Remark \ref{remark_PRB}
		\ENDIF
	}
\end{algorithmic}
\end{algorithm}
\vspace{-5pt}
\subsection{Sensing Agent Selection with Extended Kalman Filter}
Due to the complexity of sensor selection based on VoI, we adopt a heuristic approach with primary concept guiding the resolution of ~\eqref{glob_qos} is to ensure that, during each QI $t$ the minimum necessary number of $\SA$s is selected for transmission. 
This selection aims to maintain the desired level of certainty in the estimation of the state $\bs_t$, while taking into account the constraints about  VoI-based $\SA$s Scheduling and Power Control and  confidence of estimated state \eqref{scheduling_problemb}. Our approach will prioritize the former first, ensuring that the DT model keeping the freshness of physical world before addressing the latter.

The expression for the Minimum Mean Square Error (MMSE) estimator applied to a KF  is provided in \cite[Eq. (1)]{Kalman}. In this context, for aiming to minimize the (weighted) variance of state components, we employ the Extended Kalman estimator, which is common in the IoT literature~\cite{huang2019epkf}. It is important to emphasize that the Kalman equations constitute a linearization of the actual nonlinear system dynamics, thus introducing the possibility of supplementary errors.  We also assume that the virtual environment possessing complete awareness regarding the process statistics, encompassing the update function $f(\mathbf{s})$ as well as the noise covariance matrices. Such an assumption is commonly adopted within the relevant literature, as the estimation of system statistics can be feasibly accomplished prior to deployment.  The primary strategy to solve \eqref{scheduling_problem} involves selecting the minimum number of $\SA$s to transmit at each QI $t$ in order to maintain the required level of estimation certainty for state $\bs_t$ as specified by $\{\bar{\xi}_{t,k}\}$. Steps of our proposed algorithm are listed in Algorithm~\ref{scheduling_alg}, effectively addresses problem \eqref{scheduling_problem}.

In the dynamic scenario, the initial state $\bs_t$ is considered a random vector characterized by a specific mean $\mathbb{E}[\bs_t] = \boldsymbol{\mu}_{\bs_0}$ and covariance matrix $\text{Cov}[\bs_0]= \mathbf{C}{\bs_0}$.  $\Qcal_t$ is initialized as an empty set due to the absence of any prior information. The EKF then calculates the estimation errors for the belief 
\begin{align}
	\hat{\bs}_t \sim \mathcal{CN}(\bmu_{\hat{\bs}_t},\bPsi^{\mathtt{pr}}_t)
\end{align}
at the $\PA$ based on prior updates $\hat{\bs}_{t-1}$ as described by 
\begin{align}\label{prior_error}
	\bPsi^{\mathtt{pr}}_{\hat{\bs}_t} = \bP\bPsi_{\hat{\bs}_{t-1}}\bP^\top + \bC_{\bu_t},
\end{align}
where the Jacobian matrix $\bP= \mathcal{J}\{f(\bs_{t-1})\}$  linearizing the nonlinear model of $f(\bs_{t-1})$. At the beginning, to guarantee the AoL condition \eqref{scheduling_problemd} has been fulfilled, any feature $s_{t,k}$ fails to satisfy \eqref{scheduling_problemd}, we proceed to get the corresponding $\SA_m$ to emprove AoL status as
\begin{align}\label{AoL_update}
    \SA_{m}^{*} = \{\underset{\SA_{m} \in\mathcal{P}_t}{\ \mathrm{argmin }\ }
d_{\SA_m, \PA} |\SA_{m} \rightarrow s_{t,k}, m\in\Mcal\},
\end{align}
where $d_{\SA_m, \PA}$ indicates the distance from $\SA_m$ to $\PA$. We then update the schedule $\SA$ set $\Qcal_t$ by adoptting 
\begin{equation}\label{update_set}
    \Qcal_t \leftarrow \Qcal_t \cup\{ \SA_m^* \};\  \Pcal_t \leftarrow\Pcal_t\backslash\{\SA_{{m}}^*\}.
\end{equation}
Following the specification of a given set of error variance qualities $\{\bar{\xi}_{t,k}\}_{t\in\Tcal, k\in\Kcal}$, the conditions specified in \eqref{glob_qos} and \eqref{scheduling_problemd} yield two potential outcomes:
\begin{enumerate}
	\item In case that the those conditions are satisfied for all $k\in\Kcal$, the DT model achieves the required confidence and AoL bound without requiring any observation from the $\SA$s. The prior update alone suffices, resulting in an empty set for $\Qcal^*_t = \emptyset$.
	\item If any of conditions is violated, indicating that at least one noteworthy feature lacks sufficient accuracy in its estimation, the acquisition of the corresponding observations becomes essential to enhance the estimation process, as dictated by the scheduling approach implemented in our proposed heuristic.
\end{enumerate}
It is obvious that in the first  scenario, the computation of the belief $\hat{\bs}_t$ can be achieved through the EKF blind update operation as follows:
\begin{align}\label{mu_prior}
	\hat{\bs}_t = {\hat{\bs}^{\mathtt{pr}}_t}  =f(\hat{\bs}_{t-1}) + \bu_t.
\end{align}
In the second case, Algorithm~\ref{scheduling_alg} is utilized to identify the $\SA$s with the highest VoI for querying their observations.
In order to identify the most suitable candidate feature $s_{t,k}^{*}$, where $k\in\Kcal$, an optimization problem is formulated at the $i$-th iteration as
\begin{subequations} \label{finding_state}
    \begin{alignat}{2}
        s_{t,k}^{*} =& \underset{s_{t,k}\in\bs_t}{\ \mathrm{argmax }} &\quad & 
        \frac{[\bPsi^{(i)}_t]_{k }}{\bar{\xi}^2_{t,k}} \\
        &\mbox{subject to} &&  \SA_{m} \in\mathcal{P}_t, \forall m\in\Mcal, \label{finding_stateb}\\
        &&& \SA_{m} \rightarrow s_{t,k}, \forall m\in\Mcal,
    \end{alignat}
\end{subequations}
where the notation $\SA_{m} \rightarrow s_{t,k}$ signifies that the $\SA_m$ measures feature $s_{t,k}$. We note that  at the $i$-th iteration, if any constraint in \eqref{glob_qos} is still unmet and $	|\Qcal_t| <\ C,\ |\mathcal{P}_t| >0$, 
there is room for scheduling new $\SA$s to join $\Qcal_t$. 
According to constraint \eqref{finding_stateb}, feature $s_{t,k}^{*}$ is selected only if at least one $\SA_{{m}}\in \Pcal_t$ can provide coordinating observations. Then,  $\SA_m^*\in\mathcal{P}_t$ measuring feature $s_{t,k}^{*}$ with the minimum error covariance is chosen to send its measurement. The scheduled and available $\SA$ sets $\Qcal_t$ and $\mathcal{P}_t$ can be updated by \eqref{update_set}.
$\bH_t$ and $\bC_{\bw_t}$ are the combination observation and covariance matrices, which are respectively formulated as
\begin{align}
	\bH_t &= [\bH_{1};\bH_2;\dots;\bH_{|\Qcal_t|}], \label{H_compute} \\
	\bC_{\bw_t} &=  \text{diag}[\bC_{\bw_1}, \bC_{\bw_2},\dots,\bC_{\bw_{|\Qcal_t|}}]\label{Cw_compute},
\end{align}
where $\bH_{{m}}$ is the  observation matrix of  the $\SA$ ${{m}} ({{m}}\in\Qcal_t)$. From here, we can compute the EFK gain by
\begin{align}
	\bK_t =  \bPsi^{\mathtt{pr}}_{\hat{\bs}_t}{\bH}_t^\top\big(\bC_{\bw_t}  + {\bH}_t \bPsi^{\mathtt{pr}}_{\hat{\bs}_t}{\bH}_t^\top\big)^{-1},
\end{align}
The posterior error covariance matrix is derived by
\begin{align}\label{pos_variance}
	\bPsi^{\mathtt{pos}}_{\hat{\bs}_t}=  (\bI - \bK_t\bH_t) \bPsi^{\mathtt{pr}}_{\hat{\bs}_t},
\end{align}
The iterative loop continues as long as all three of the following conditions are true: 
\begin{align}
	\left\{\begin{matrix*}[l]\label{checking3_conditions}
		\{|\Qcal^*_t| &< C;\\ 
		\exists\ [\bPsi^{\mathtt{pos}}_{\hat{\bs}_t}]_{k} &\geq \bar{\xi}^2_{k}; \\ 
		\exists\ s_{t,k}^{*} &\mbox{ as a solution in }\eqref{finding_state}\}.
	\end{matrix*}\right.
\end{align}
Hence, it makes intuitive see that the loop repeats for at maximum $C$ iterations before terminating. 
The posterior update is then  mathematically expressed by
\begin{equation}\label{pos_update}
	{\hat{\bs}^{\mathtt{pos}}_t}  = {\hat{\bs}^{\mathtt{pr}}_t} + \bK_t(\bo_t - \bH_t{\hat{\bs}^{\mathtt{pr}}_t} ),
\end{equation}
where $\bo_t $ represents the combination of received $\SA$ observations. Accordingly, we update $\hat{\bs}_t = \hat{\bs}^{\mathtt{pos}}_t$. Despite the local $\SA$ scheduling solution, our  approach ensures the long-term balance between state certainty and communication cost over different QIs with respect to the $\PA$'s requirements.

\begin{figure*}[t]
    \begin{minipage}{0.245\textwidth}
        %		\centering
        \includegraphics[trim=.cm -1.5cm 0.cm 0cm, clip=true, width=1.8 in]{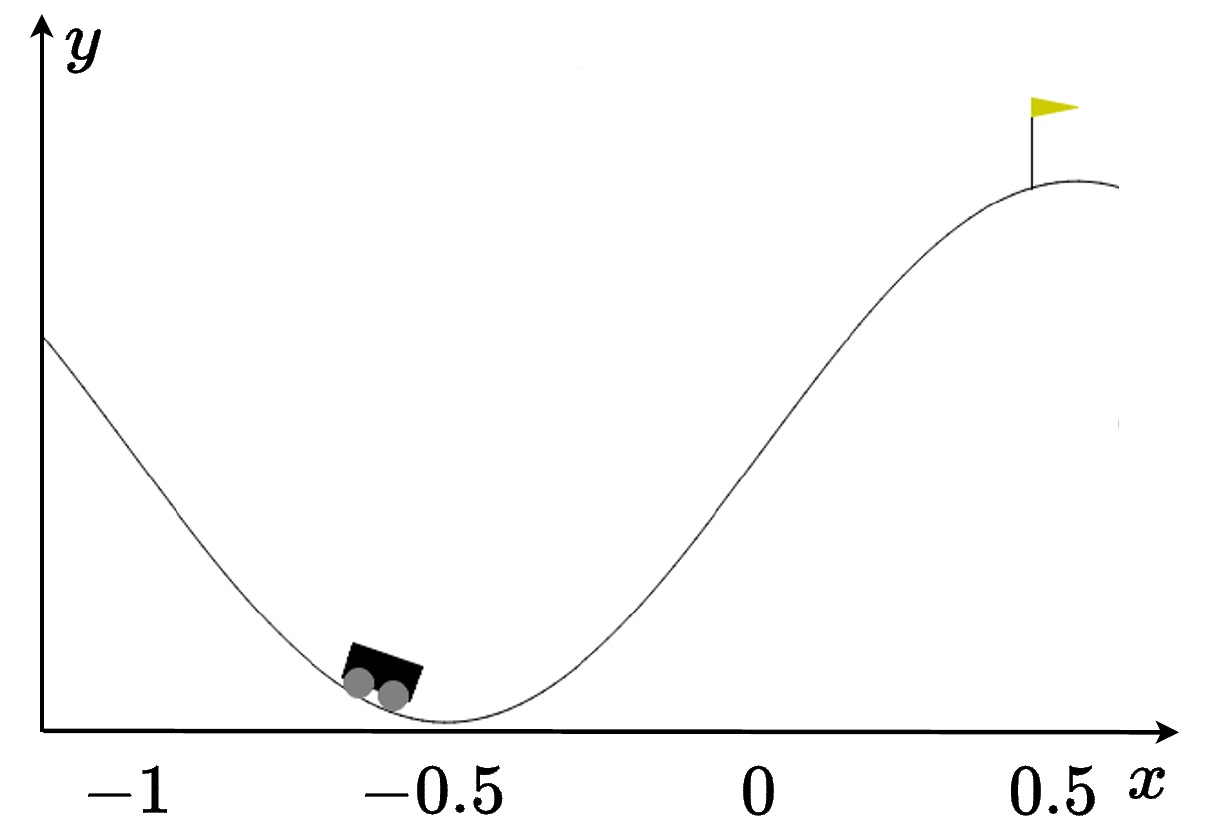} \\ 
        % \caption{heading}
        \vspace*{-13pt}
        \centering {\footnotesize$(a)$  MountainCarContinuousv0 \cite{towers_gymnasium_2023}}
        \vspace*{-0pt}
    \end{minipage}
    \begin{minipage}{0.245\textwidth}
        %		\centering
        \includegraphics[trim=0.cm 0cm 0.cm 0cm, clip=true, width=1.8 in]{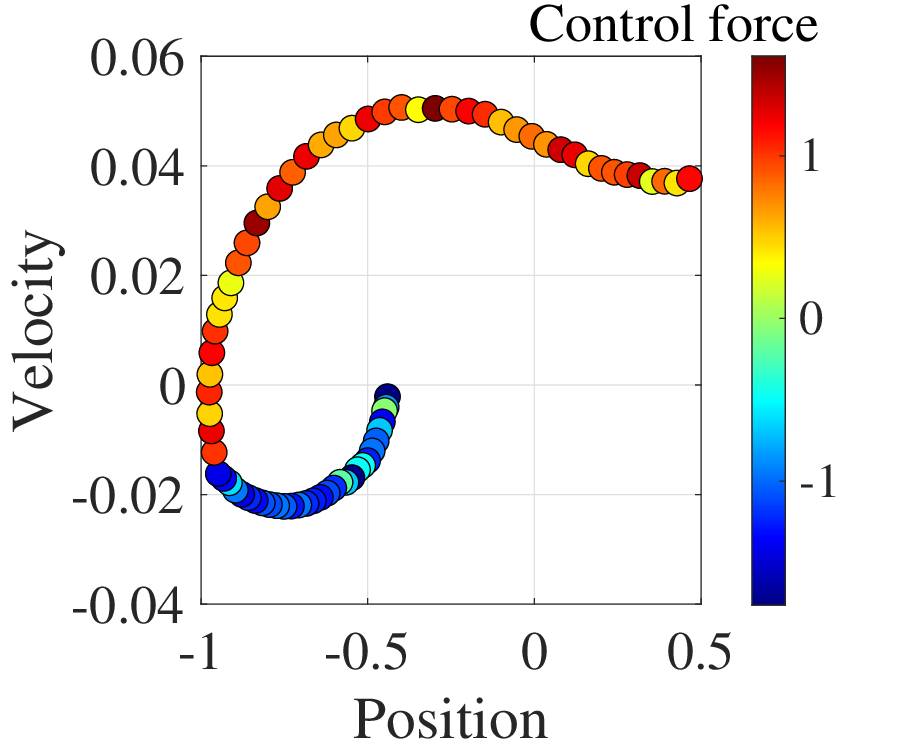} \\ 
        \vspace*{-0pt}
        \centering {\footnotesize$(a)$  AoL-REVERB (75 QIs)}
        \vspace*{-0pt}
    \end{minipage}
    \begin{minipage}{0.245\textwidth}
        %		\centering
        \includegraphics[trim=0.cm 0cm 0.cm 0cm, clip=true, width=1.8 in]{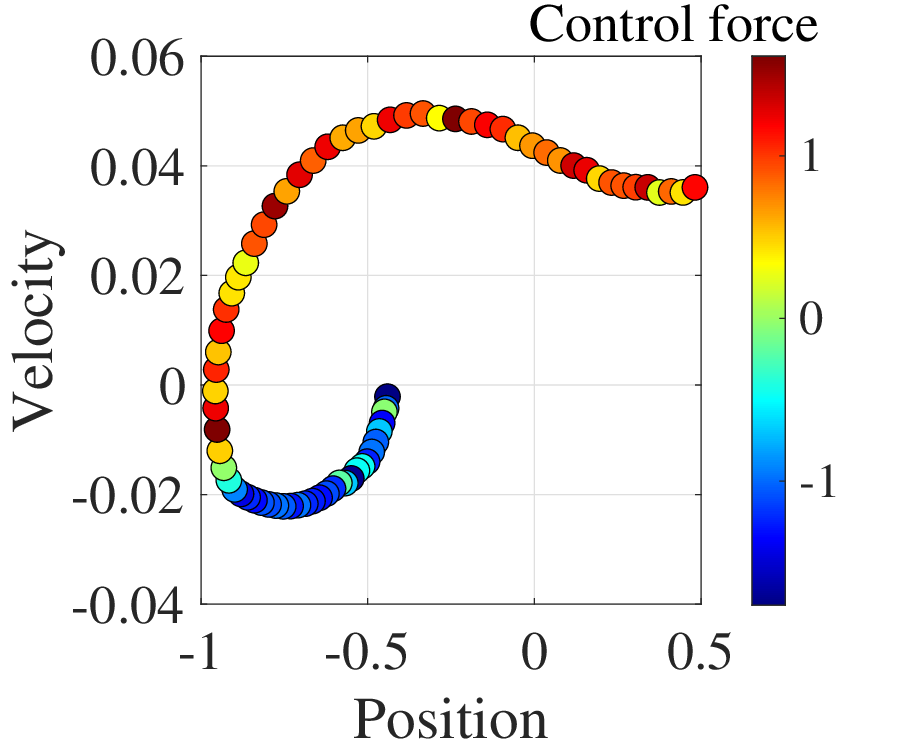} \\ 
        \vspace*{-0pt}
        \centering {\footnotesize$(b)$  Perfect  (75 QIs) }
        \vspace*{-0pt}
    \end{minipage}
    \begin{minipage}{0.245\textwidth}
        %		\centering
        \includegraphics[trim=0cm 0cm 0.cm 0cm, clip=true, width=1.8 in]{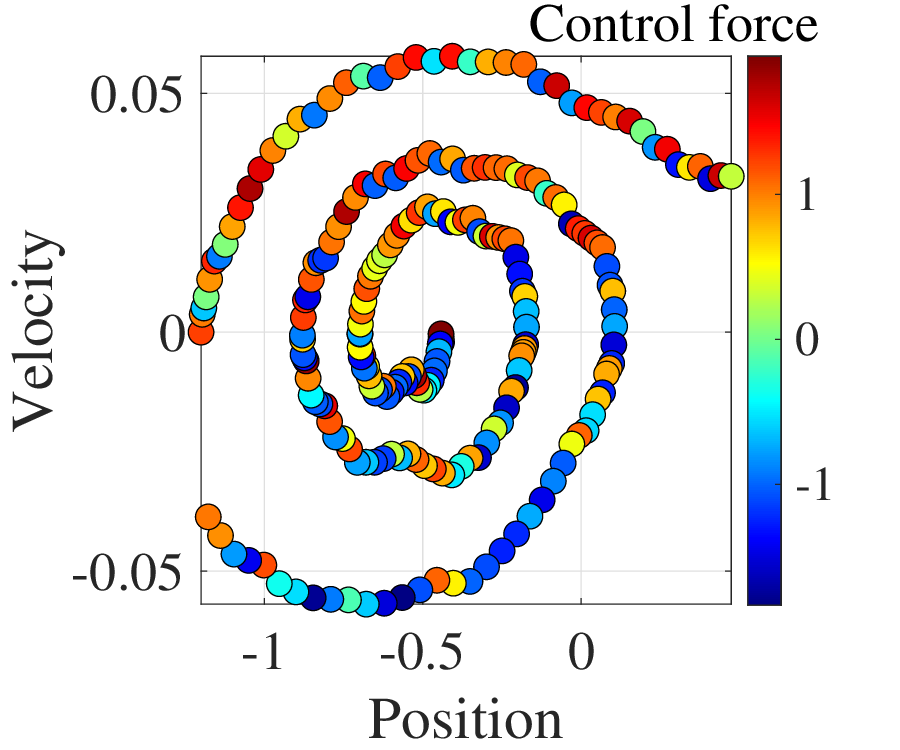} \\ 
        \vspace*{-0pt}
        \centering {\footnotesize$(c)$  Traditional (143 QIs)}
        \vspace*{-0pt}
    \end{minipage}
    \caption{The environment and trajectory with coordinated control force of different schemes.}
    \label{fig_trajectory}
    \vspace*{-15pt}
\end{figure*}
\vspace{-5pt}
\subsection{Optimal Physical Resource Block Scheduling For Uplink Transmission}
Both DL and UL transmissions are subject to potential latency during data delivery, which is influenced by the channel quality and the scheduled transmission resource. 
The channel between $\SA_m$ and AP is modeled as Rician channel with strong LoS link and small-scale fading with rich scattering, where the instantaneous Signal-to-Noise (SNR) ratio is modeled as
\begin{align}\label{SNR}
    \gamma_{t,m} = \frac{\Gamma p_{t,m}^\mathrm{tx}}{d^{\alpha}_mWN_0} \mathcal{G}_m, \forall m \in \Mcal,
\end{align}
with  $\Gamma$ is constant depending on the system parameter (operating frequency and antenna gain); $p_{t,m}^\mathrm{tx}$ is the transmitted power, $d_m$ represents the distance between device $m$ and AP. $\alpha$ is the path loss exponent; $W$ is the allocated bandwidth and $N_0$ stands for noise power. $\mathcal{G}_m$ represents the fading power with expected value $\bar{\mathcal{G}} \triangleq \mathbb{E}[|\mathcal{G}_m|^2] =1$. Herein, we adopt $\mathcal{G}_m$ as Rician distribution with a non-central
chi-square probability distribution function (PDF) which is expressed as \cite{simon2008digital}
\begin{align}
    f_\mathcal{G}(w) = \frac{(G+1)e^{-G}}{\bar{\mathcal{G}}}e^\frac{-(G+1)w}{\bar{\mathcal{G}}}I_0\Big(2\sqrt{\frac{G(G+1)w}{\bar{\mathcal{G}}}}\Big),
\end{align}
where $w\geq 0$ and $I_0(\cdot)$ denotes the zero-order modified Bessel function of the first kind, while $G$ represents the Rician factor, signifying the ratio between the power within the Line-of-Sight (LoS) component and the power distributed among the non-LoS multipath scatters. 
Assuming CSI is known at sender/receiver, we use Shannon’s bound to represent the rate $R_{t,m}$ of every link as %\pp{PP: To justify this, we need to assume that the sender/receiver have the CSI.}
\begin{align}
	R_{t,m} = W\log_2(1+\gamma_{t,m}).
\end{align}
% $R_{t,m} = W\log_2(1+\gamma_{t,m})$. 
The optimal transmission bandwidth  for each scheduled $\SA$ is computed through the following lemma.
\begin{lemma}\label{optimal_bandwidth}
    The optimal transmission bandwidth $W$  to satisfy the reliability of outage perfromance \eqref{aoi_requirement} is  expressed in closed form as
    \begin{equation}\label{BW_require}
        W^*_{t,m} = \frac{-D\log(2)}{\tau^\mathrm{max}\left(\mathbb{W}\left(-\frac{e^{-{1}/{\Theta}}}{\Theta}\right) + \frac{1}{\Theta}\right)}.
    \end{equation}
    where 
	\begin{IEEEeqnarray}{lll}
		\Theta &= \frac{\Gamma p_{t,m}^\mathrm{tx}y_\tau^2\tau^\mathrm{max}}{2(1+G)d^{\alpha}_mN_0D\log(2)}, \non\\
		y_\tau &= \sqrt{2G}+\frac{1}{2Q^{-1}(\varepsilon)}\log(\frac{\sqrt{2G}}{\sqrt{2G}-Q^{-1}(\varepsilon)})-Q^{-1}(\varepsilon), \non
	\end{IEEEeqnarray}
	% $y_\tau = \sqrt{2G}+\frac{1}{2Q^{-1}(\varepsilon)}\log(\frac{\sqrt{2G}}{\sqrt{2G}-Q^{-1}(\varepsilon)})-Q^{-1}(\varepsilon)$, $\Theta = \frac{\Gamma p_{t,m}^\mathrm{tx}y_\tau^2\tau^\mathrm{max}}{2d^{\alpha}_mN_0D\log(2)}$, 
	with $D$ as the length of data packet, and $Q^{-1}(\cdot)$ denoting the  inverse Q-function.
    \begin{proof}

Then, the outage probability \eqref{aoi_requirement} is  formulated as
\begin{align}
	\mathbb{P}[\Delta _{t,m} \geq  \tau^\mathrm{max}] &=  \mathbb{P}[W\log_2(1+ \gamma_{t,m}) \leq  \frac{D}{\tau^\mathrm{max}}] \\
	&=  \mathbb{P}[ \gamma_{t,m} \leq  2^\frac{D}{W\tau^\mathrm{max}}-1] \\
	& \triangleq 1 - Q(x_\tau, y_\tau) \leq \varepsilon, \label{y_Q_ex}
\end{align}
where
\begin{IEEEeqnarray}{lll}
	x_\tau &=\sqrt{2G}, \label{x_tau}\\
	y_\tau &= \sqrt{2(G+1)(2^\frac{D}{W\tau^\mathrm{max}}-1)\mathcal{G}_m/\gamma_{t,m}}, \label{y_tau}
\end{IEEEeqnarray}
with $Q(x_\tau, y_\tau)$ as the the first order Marcum Q-function.  Then, \eqref{y_tau} can be rewriten as
\begin{IEEEeqnarray}{lrl}
	&  \frac{2(G+1)(2^\frac{D}{W_{t,m}\tau^\mathrm{max}}-1)\mathcal{G}_m}{y_\tau^2} &= \gamma_{t,m} \\
	\Leftrightarrow \ &  \frac{2(G+1)(2^\frac{D}{W_{t,m}\tau^\mathrm{max}}-1)\mathcal{G}_m}{y_\tau^2}  &=\frac{\Gamma p_{t,m}^\mathrm{tx}}{d^{\alpha}_mWN_0} \mathcal{G}_m\\
	\Leftrightarrow \ & \log_2\Big({1 + \frac{\Gamma p_{t,m}^\mathrm{tx}y_\tau^2}{2(1+G)W_{t,m}d^{\alpha}_mN_0}} \Big) &= \frac{D}{W_{t,m}\tau^\mathrm{max}} \\
	\Leftrightarrow \ &  {1 + \frac{\Gamma p_{t,m}^\mathrm{tx}y_\tau^2}{2(1+G)W_{t,m}d^{\alpha}_mN_0}} = \exp&\left({\frac{D\log(2)}{W_{t,m}\tau^\mathrm{max}}}\right).\label{eq_next}
\end{IEEEeqnarray}
Let us denote 
\begin{IEEEeqnarray}{lrl}\label{Upsilon_Theta}
	\Upsilon = \frac{-D\log(2)}{W_{t,m}\tau^\mathrm{max}},\ \Theta = \frac{\Gamma p_{t,m}^\mathrm{tx}y_\tau^2\tau^\mathrm{max}}{2(1+G)d^{\alpha}_mN_0D\log(2)},
\end{IEEEeqnarray}
we have the equivalent fomulation of \eqref{eq_next} as
\begin{IEEEeqnarray}{lrl}
	&1- \Upsilon\Theta  &= e^{-\Upsilon} \\
	\Leftrightarrow \ & (1-\Upsilon\Theta)e^{\Upsilon}  &= 1 \\
	\Leftrightarrow \ & -\Theta\nu e^{\nu + 1/\Theta}  &= 1, \text{ where } \nu = \Upsilon - 1/\Theta \\
	\Leftrightarrow \ & \nu e^{\nu}  &= -\frac{e^{-{1}/{\Theta}}}{\Theta} \label{eq_next_2}
\end{IEEEeqnarray}
Apply the product logarithm to \eqref{eq_next_2}, we can achieve
\begin{IEEEeqnarray}{lll}
	\nu = \mathbb{W}\left(-\frac{e^{-{1}/{\Theta}}}{\Theta}\right), \label{eq_next_3}
\end{IEEEeqnarray}
where $\mathbb{W}(\cdot)$ is the Lambert W function. Plug \eqref{eq_next_3} into $\nu = \Upsilon - 1/\Theta $, we obtain
\begin{IEEEeqnarray}{lll}
	\Upsilon = \mathbb{W}\left(-\frac{e^{-{1}/{\Theta}}}{\Theta}\right) + \frac{1}{\Theta}. \label{eq_next_4}
\end{IEEEeqnarray}
Using this result and the definition of $\Upsilon$ in \eqref{Upsilon_Theta}, we can derive the optimal transmission bandwidth $W_{t,m}$ as in \eqref{BW_require}.

For finding the value of $y_\tau$, we stress that at the maximum tolerable value of $\varepsilon$, then $y_\tau = Q^{-1}(x_\tau, 1-\varepsilon)$ formed as the inverse Marcum Q–function respecting to second argument. In this study, we consider a Rician channel with strong LoS component, i.e., $G>G_0^2/2$ and $Q^{-1}(\varepsilon) \neq 0$, where $G_0$ is  the intersection of the sub-functions at $x_\tau > \max[0, Q^{-1}(\varepsilon)]$. 

Therefore, the approximated form of $y_\tau$ as \cite[Lemma 1]{8017572}:
\begin{IEEEeqnarray}{lll}
	y_\tau = \sqrt{2G}+ \frac{1}{2Q^{-1}(\varepsilon)}\log\Big(\frac{\sqrt{2G}}{\sqrt{2G}-Q^{-1}(\varepsilon)}\Big) - Q^{-1}(\varepsilon).\nonumber\\ \label{y_tau_final}
\end{IEEEeqnarray} 
We complete the proof.
\end{proof}

%	\vspace{-10pt}
\end{lemma}
\begin{table}[H]
\vspace*{-15pt}
\centering
\caption{Simulation Parameters}
\resizebox{8cm}{!} 
{
    \begin{tabular}{ll}
        \hline
        Parameter & Value \\
        \hline
        Carrier frequency ($f_c$)          & 2.4 GHz      \\
        Power budget at $\{\SA_m\}$   (P)       & 20 mW      \\
        DT required error variances $(\xi^2_{pos}, \xi^2_{vel})$ & (0.01, 0.002)     \\
        Max. connection ($C$) & 10 \\
        Rician factor ($G$) & 10 dB \\
        System parameter ($\Gamma$) & 1 \\
        Maximum latency ($\tau^\mathrm{max}$)          &   5 ms    \\
        Channel noise power          & -11.5 dBm     \\
        Max. distance ($d^{\max}$)          & 20 m     \\
        Leaning rate (both actor and critic) & $1e^{-3}$ \\
        Outage probability factor $(\varepsilon)$&  $1e^{-5}$\\
        Data packet size $(D)$&  $1024$ bits\\
        \hline
    \end{tabular}
}
\label{parameters}
\end{table}
\begin{remark}\label{remark_PRB}
Using Lemma~\ref{optimal_bandwidth}, we can compute the coordinated required bandwidth for each allocated $\SA_m$. Then, we evaluate the  uplink behavior considering the 3GPP 4-bit Table 7.2.3-1 \cite{3GPP_PRB} the DT can schedule the accordinated number of PRBs to $\SA_m$ at each QI $t$, which has been widely adopted in the literature and the reference therein~\cite{de2020radio, WirelessSuite}.
\end{remark}

\begin{figure*}[t]
    \begin{minipage}{0.33\textwidth}
        %		\centering
        \vspace*{5pt}
        \includegraphics[trim=0cm 0cm 0.cm 0cm, clip=true, width=2 in]{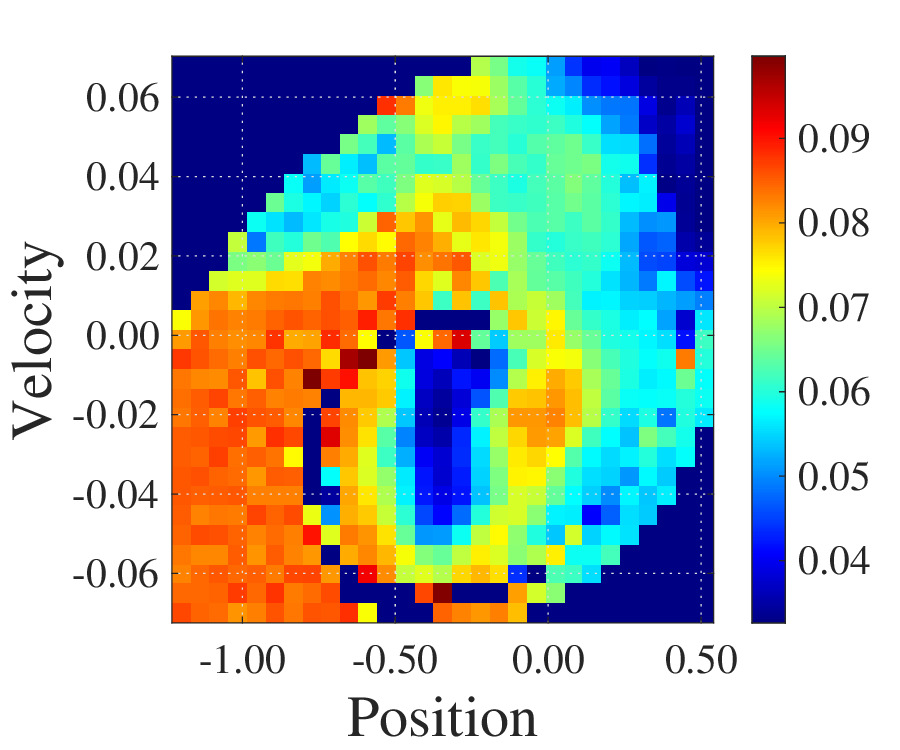} \\ 
        \vspace*{-0pt}
        \centering {\footnotesize$(a)$  Position uncertainty}
        \vspace*{-0pt}
    \end{minipage}
    \begin{minipage}{0.33\textwidth}
        %		\centering
        \includegraphics[trim=0.cm 0cm 0.cm 0cm, clip=true, width=2 in]{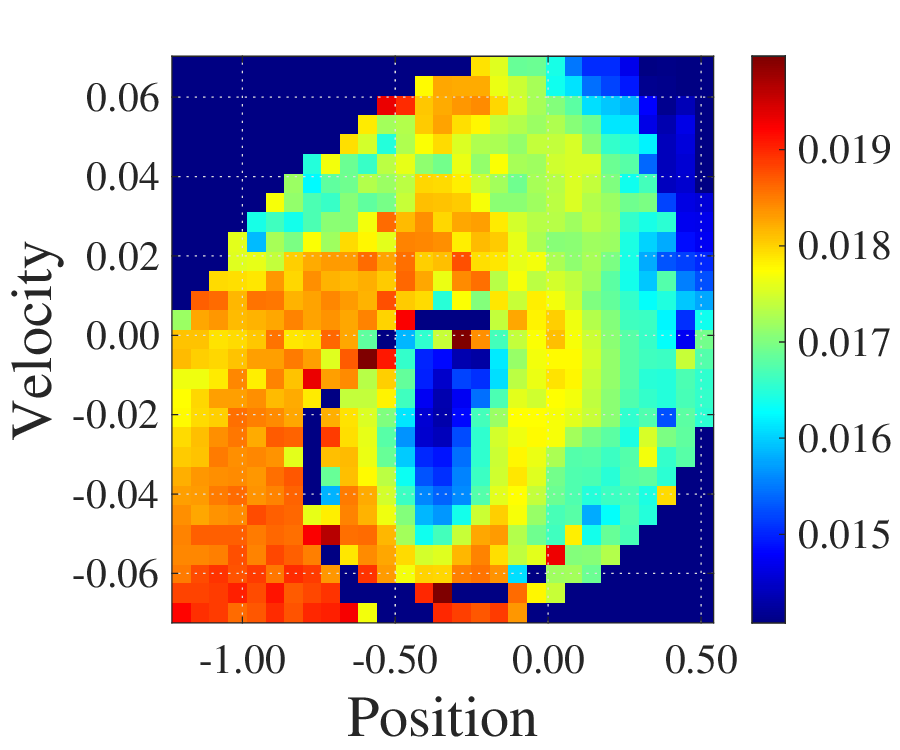} \\ 
        \vspace*{-5pt}
        \centering {\footnotesize$(b)$  Velocity uncertainty}
        \vspace*{-0pt}
    \end{minipage}
    \begin{minipage}{0.33\textwidth}
        %		\centering
        \includegraphics[trim=0cm 0cm 0.cm 0cm, clip=true, width=2 in]{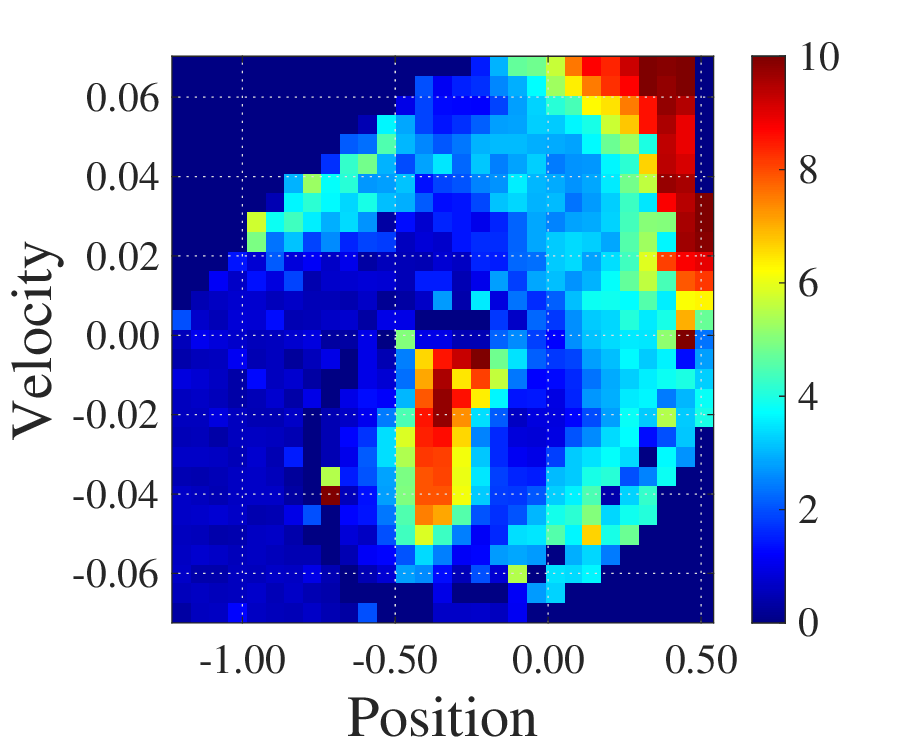} \\ 
        \vspace*{-5pt}
        \centering {\footnotesize$(c)$  Number of selected $\SA$s}
        \vspace*{-0pt}
    \end{minipage}
    \caption{Performance evaluation of AoL-REVERB solution:  $(a)$ and $(b)$ are position and velocity uncertainty level ($\{\bar{\xi}_{t,k}\}$);  $(c)$ represents  number of selected $\SA$s under position-velocity coordinate.}
    \label{AoL-REVERB_performance}
    \vspace*{-15pt}
\end{figure*}
\begin{figure}[t]
    \begin{minipage}{0.45\textwidth}
        \centering
        \includegraphics[trim=0.cm 0cm 0cm 0cm, clip=true, width=1\textwidth]{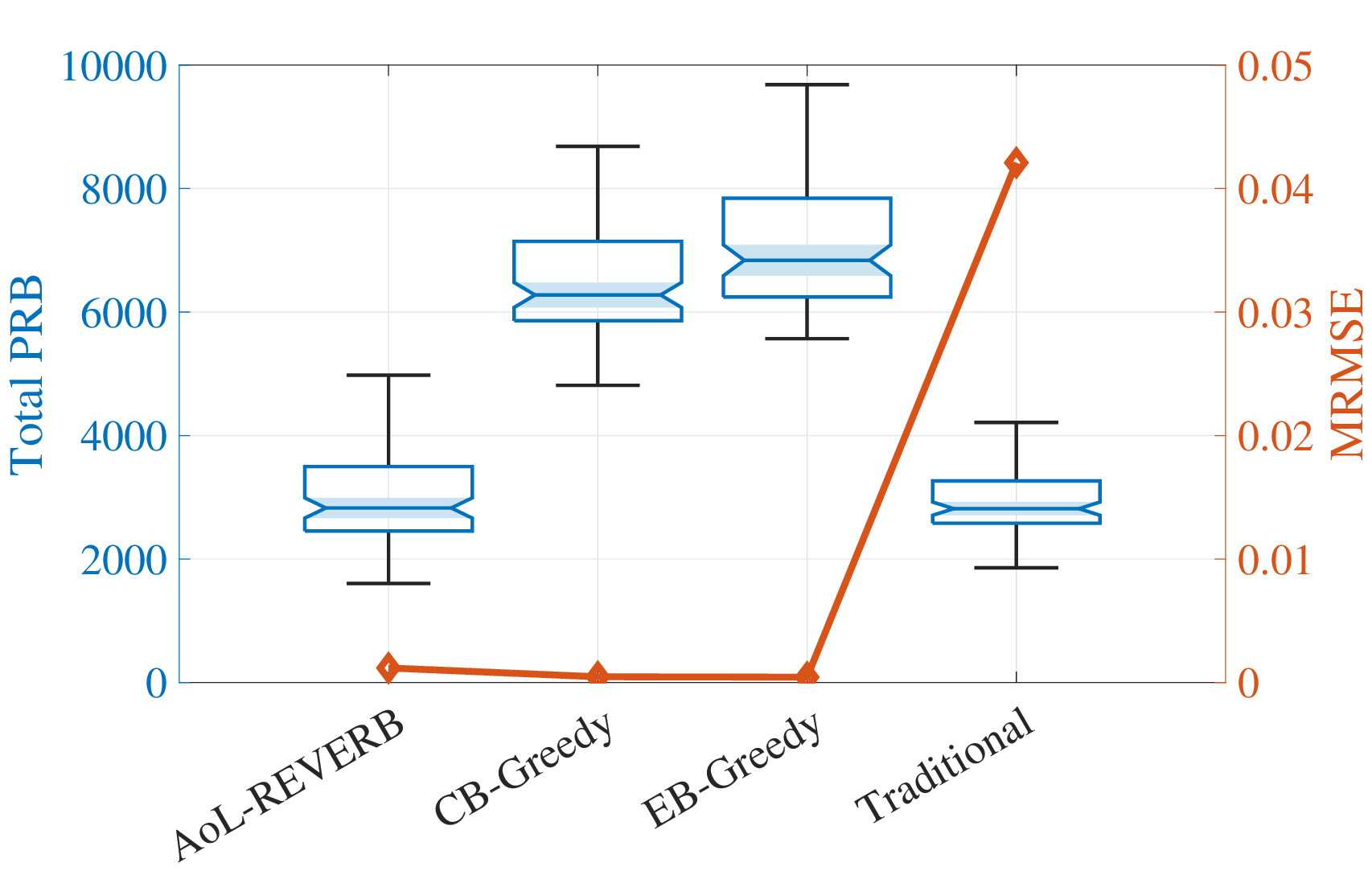} \\ 
        \caption{Consumed PRB \& MRMSE of various schemes.}
        \label{fig_PRBvsMRMSE}
    \end{minipage}
    \vspace*{-0pt}
    \begin{minipage}{0.45\textwidth}
        \centering
        \includegraphics[trim=0.cm 0cm 0cm 0cm, clip=true, width=1\textwidth]{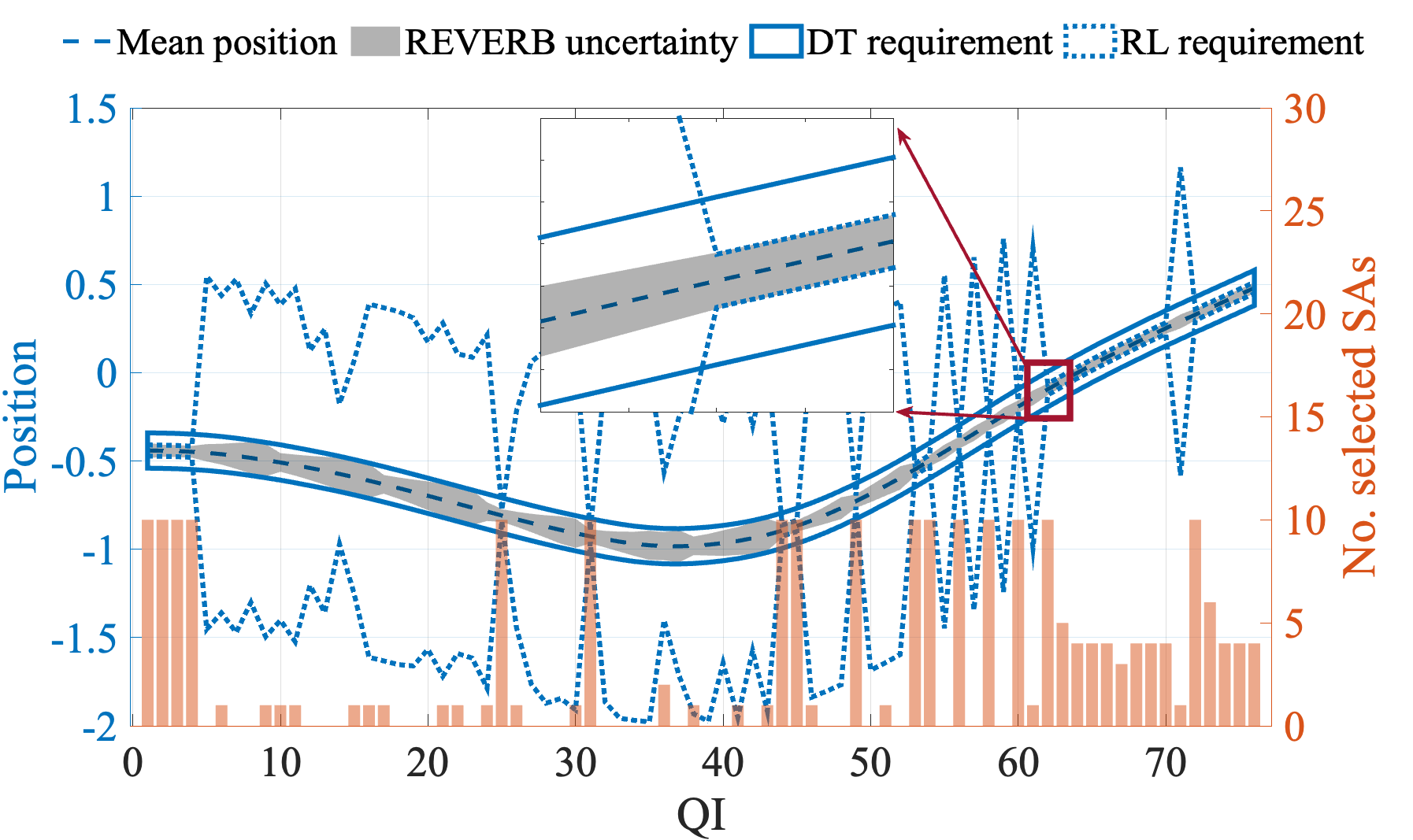} \\ 
        \captionsetup{format=plain, justification=justified, width=1\linewidth}
        \caption{ Snapshot of AoL-REVERB uncertainty evolution and number of  selected $\SA$s vs QI to meet both the uncertainty requirements of DT and control solution.}
        \label{fig_uncertainty_evolution}
    \end{minipage}
    \vspace*{-15pt}
\end{figure}
\vspace*{-10pt}
%%%%%%%%%%%%%%%%%%%%%%%%%%%%%%%%%%%%%%%%%%%%%%%%
\section{Numerical Results}
%%%%%%%%%%%%%%%%%%%%%%%%%%%%%%%%%%%%%%%%%%%%%%%%

To simultaneously investigate our proposed framework encompassing system dynamics and control, we conducted an examination using our AoL-REVERB algorithm within the MountainCarContinuousv0 environment provided by OpenAI Gym~\cite{moore1990efficient}. The state vector $\bs_{t} = [x_t, \dot{x_t}]^\top$ includes position $x_t$ and velocity $\dot{x_t}$. The observation matrices are given by $\bH_{pos} = [1,  0]; \text{ and } \bH_{vel} = [0,  1]$,  
respectively for the position and velocity. Other important parameters are included in Table~\ref{parameters}. In the DT model, the agent makes decisions concerning the applied force $a_t\in[-1,1]$ on the car and selects an accuracy level denoted by $\boldsymbol{\eta}_t = [\eta_{t,1}, \eta_{t,2}]^\top$. The original reward for each QI in the environment is represented as 
$ {r}_t = -0.1 \times a_t^2.$ 
In our system, we adopt a modified reward as indicated in \eqref{reward_shaping}:
\begin{align}\label{reward_shaping1}
\hat{r}_t = {r}_t + \kappa \times \Big(\frac{1}{2}\sum_{i =1}^{2}\eta_{t,i}\Big),
\end{align}
where the positive weighted parameter $\kappa>0$ is set to $\kappa = 5\times 10^{-6}$. Additionally, the original environment includes a termination reward, which the agent receives when the car successfully reaches the target position at 0.45. This termination reward is also provided to our agent upon successful completion. The evolution of the state in \eqref{dynamic_model} is  defined with
\begin{align}
f(\bs_{t-1})&= \begin{bmatrix}
	\dot{x}_{t-1}	\\ 
	-\varphi \cos(3x_{t-1})	
\end{bmatrix},\quad 
\bB &= [0, \vartheta]^\top,
\end{align}
where the constants $\varphi = 0.0025$ and $\vartheta = 0.0015$. $M$ $\SA$s are assuming placed randomly within an area where the maximum distance $d^\mathrm{max}$ from $\SA$s to AP.

We compare our proposed REVERB and four other benchmarks: \textit{Perfect} allows DT to get the observation of the next state without any error; Cost-based greedy (\textit{CB-Greedy}) indicates that the AP queries all nearest $\SA$s based on ascending order of distance from AP to $\SA$ at each QI. Error-based greedy (\textit{CB-Greedy}) \cite{10092861} is similar to \textit{CB-Greedy} but it relies on the decreasing confidence levels of $\SA$s. In the greedy benchmarks is $|\Qcal_t^*| = C$. We also consider the \textit{Traditional} scheme~\cite{sutton2018reinforcement} where the RL agent gets noise observation from one $\SA$ every QI without Algorithm 2. All schemes are conducted under 1000 Monte Carlo simulations.

In Fig.~\ref{fig_trajectory}, we compare the trajectory evolution with coordinated force of REVERB, \textit{Perfect}, and \textit{Traditional} schemes. It is observed that AoL-REVERB's trajectory is same performance to the \textit{Perfect} when using 75 QIs to reach the goal. On the contrary, the trained network with the \textit{Traditional}  must use up to $2\times$ number of steps to reach the goal. These findings validate the reliability of AoL-REVERB in adjusting forces within a noisy environment. Furthermore, the comparison of control forces among the algorithms confirms our discussion in Example 1, emphasizing that precise knowledge of the car's position and velocity at all times is not necessary for implementing suitable control forces.

Fig.~\ref{AoL-REVERB_performance}(a)-(c) illustrates the uncertainty levels and the number of selected $\SA$s across the entire observation space. A comparison between Fig.~\ref{AoL-REVERB_performance}(a)-(b) and Fig.~\ref{AoL-REVERB_performance}(c) reveals that DT accumulates more data as critical positions, i.e., the agent approaches the target ($x_t \geq 0.45$) and at bend locations ($x_t\sim -0.5$). In scenarios where the uncertainty level is high but the car remains far from the target ($x_t < -0.5$), the necessity for additional observations to enhance control efficiency is deemed ineffective. Importantly, this high uncertainty does not compromise the performance of the original task but facilitates the conservation of communication resources.

\begin{figure*}[t]
    \begin{minipage}{0.33\textwidth}
        %		\centering
        % \vspace*{-10pt}
        \includegraphics[trim=0.cm 0cm 0cm 0cm, clip=true, width=1\textwidth]{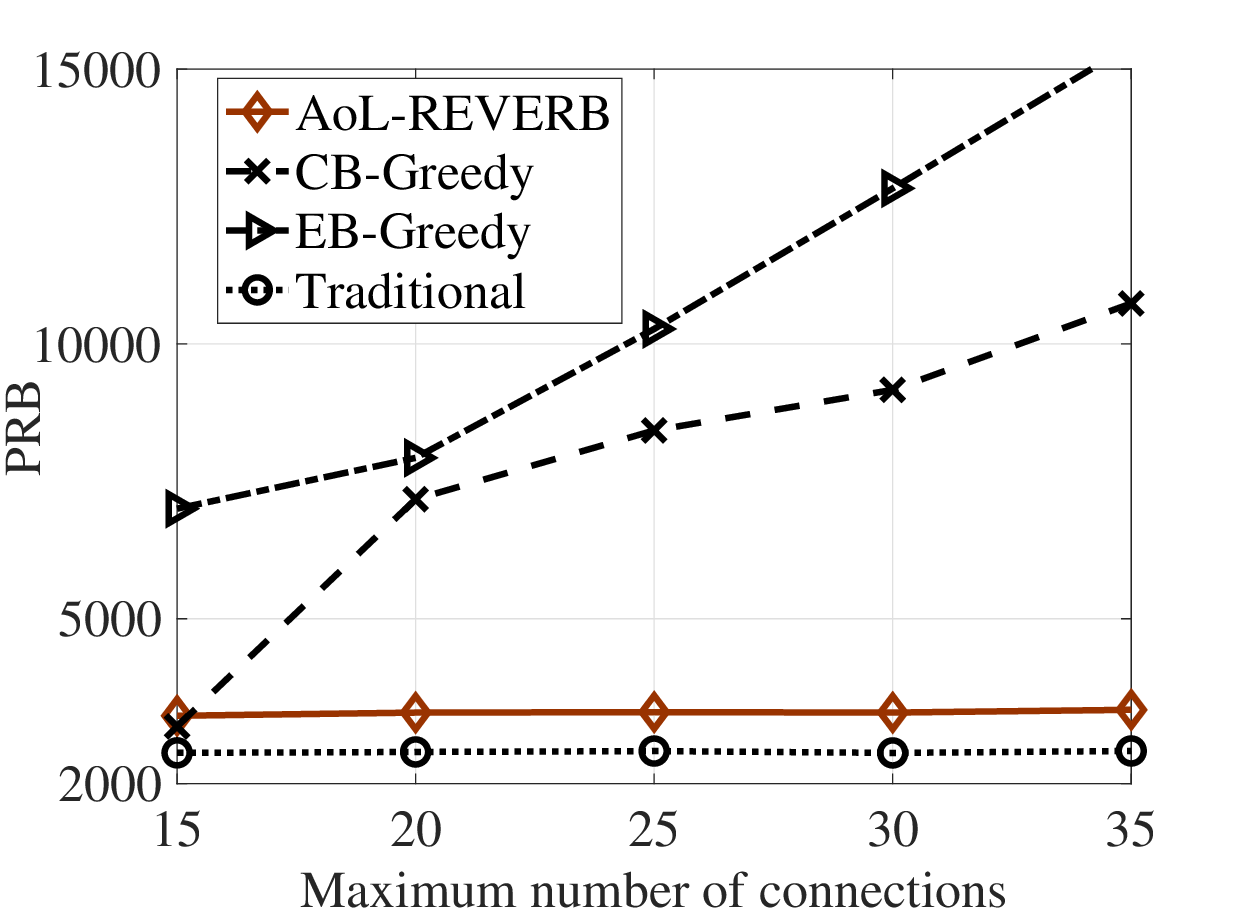} \\ 
        \centering {(a) Total PRB consumption}
        \label{}
        % \vspace*{-10pt}
    \end{minipage}
    \begin{minipage}{0.33\textwidth}
        %		\centering
        % \vspace*{-03pt}
        \includegraphics[trim=0.0cm 0cm 0.cm 0cm, clip=true, width=1 \textwidth]{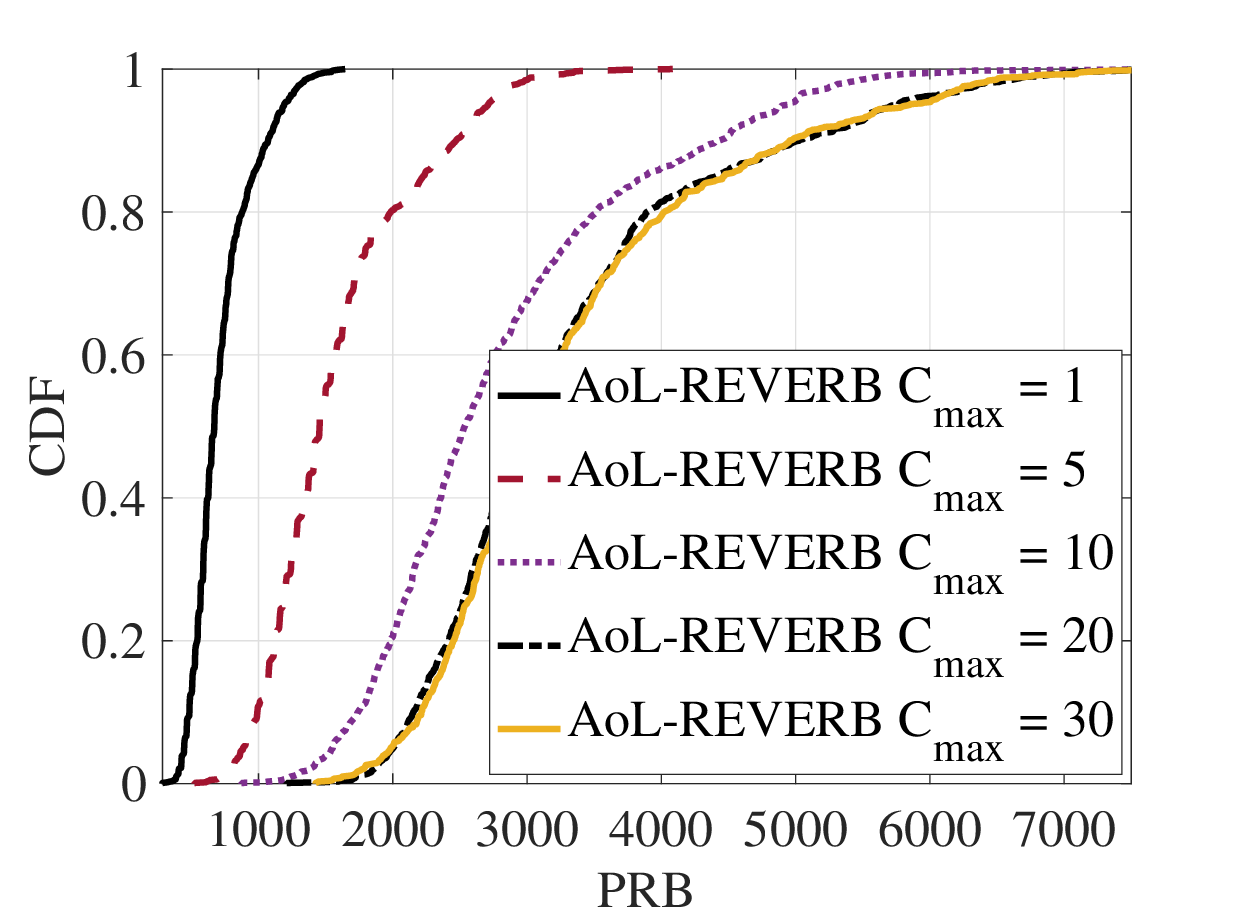} \\ 
        %		\vspace*{-15pt}
        \centering {(b) CDF of the PRB consumption} 
        \label{}
        % \vspace*{-10pt}
    \end{minipage}
    \begin{minipage}{0.33\textwidth}
        % \vspace*{-03pt}
        \includegraphics[trim=0.0cm 0cm 0.cm 0cm, clip=true, width=1 \textwidth]{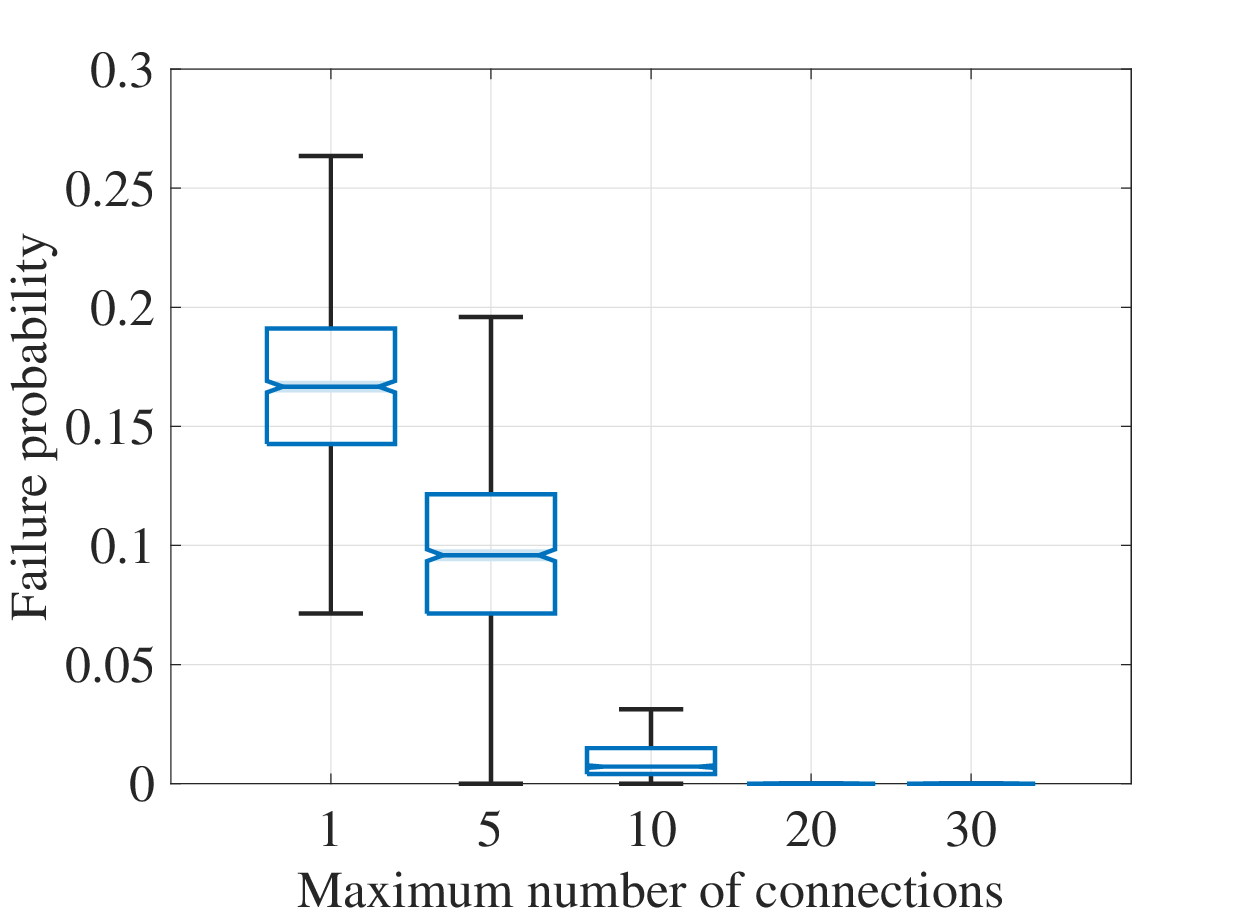} \\ 
        \centering {(c) Failure probability} 
        \label{}
        % \vspace*{-10pt}
    \end{minipage}
    %	\vspace*{-0.05cm}
        \caption{System performance under varying communication capacity.}
    \label{fig_vary_Cmax}
    \vspace{-15pt}
\end{figure*}
\begin{figure}[t]
	\begin{minipage}{0.24\textwidth}
		%		\centering
		\includegraphics[trim=0cm 0cm 0.cm 0cm, clip=true, width=1 \textwidth]{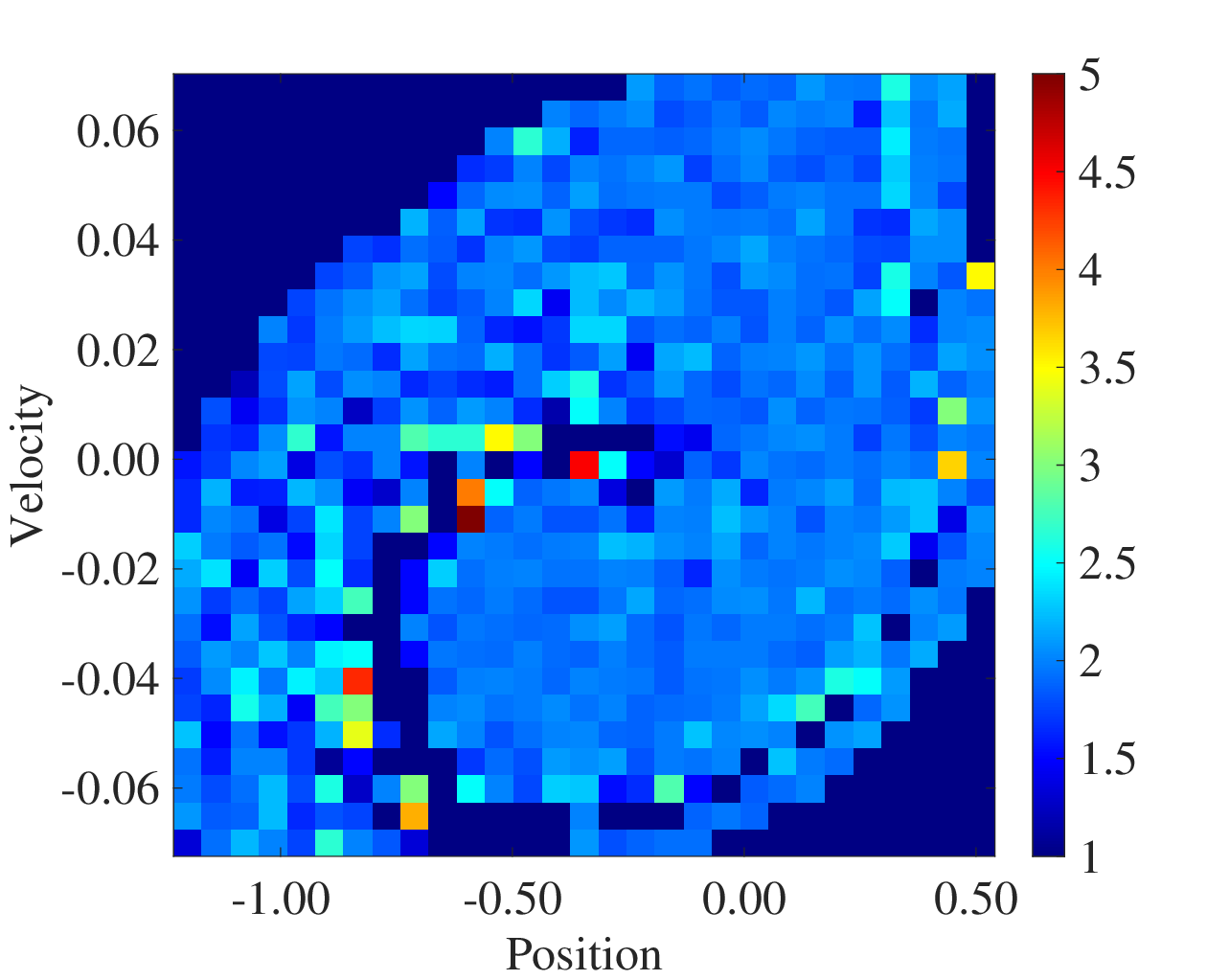} \\ 
		\vspace*{-0pt}
		\centering {\footnotesize$(a)$  Position AoL status ($\bar{\Delta}L_k = 5$)}
		\vspace*{-5pt}
	\end{minipage}
	\begin{minipage}{0.24\textwidth}
		%		\centering
		\includegraphics[trim=0cm 0cm 0.cm 0cm, clip=true, width=1 \textwidth]{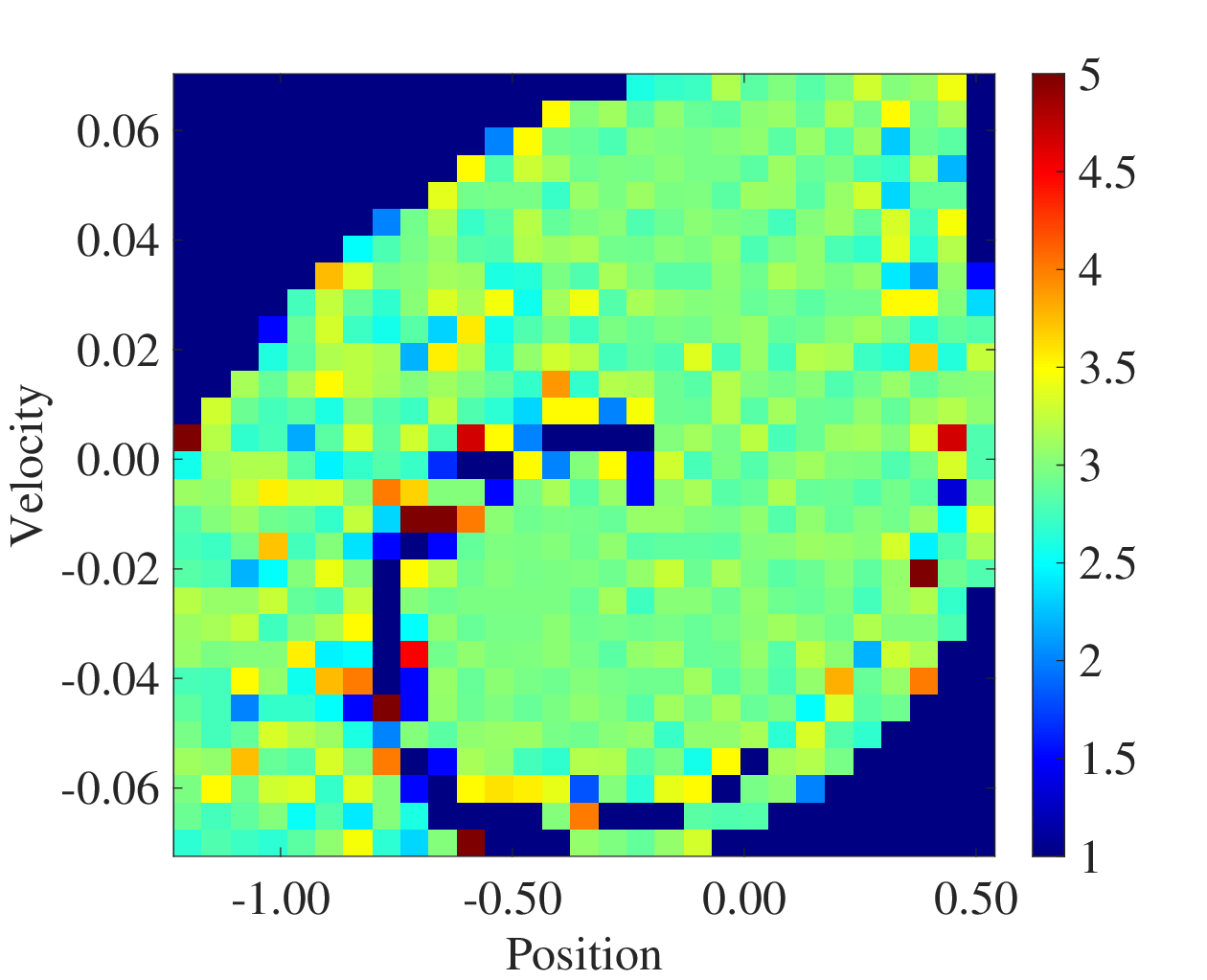} \\ 
		\vspace*{-0pt}
		\centering {\footnotesize$(b)$  Velocity AoL status ($\bar{\Delta}L_k = 5$)}
		\vspace*{-5pt}
	\end{minipage}
	\caption{AoL status under position-velocity coordinate.}
	\label{fig_AoL_status}
\vspace*{-15pt}
\end{figure}
\begin{figure}[t]
	\begin{minipage}{0.24\textwidth}
		%		\centering
		\includegraphics[trim=0cm 0cm 0.cm 0cm, clip=true, width=1 \textwidth]{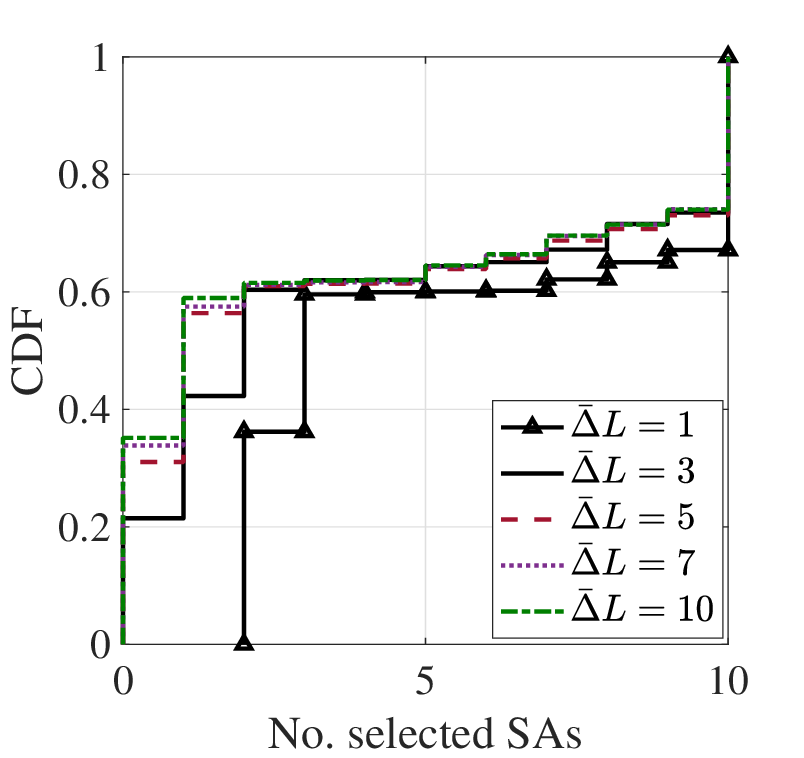} \\ 
		\vspace*{-0pt}
		\centering {\footnotesize$(a)$  CDF of the no. selected $\SA$s}
		\vspace*{-0pt}
	\end{minipage}
	\begin{minipage}{0.24\textwidth}
		%		\centering
		\includegraphics[trim=0cm 0cm 0.cm 0cm, clip=true, width=1 \textwidth]{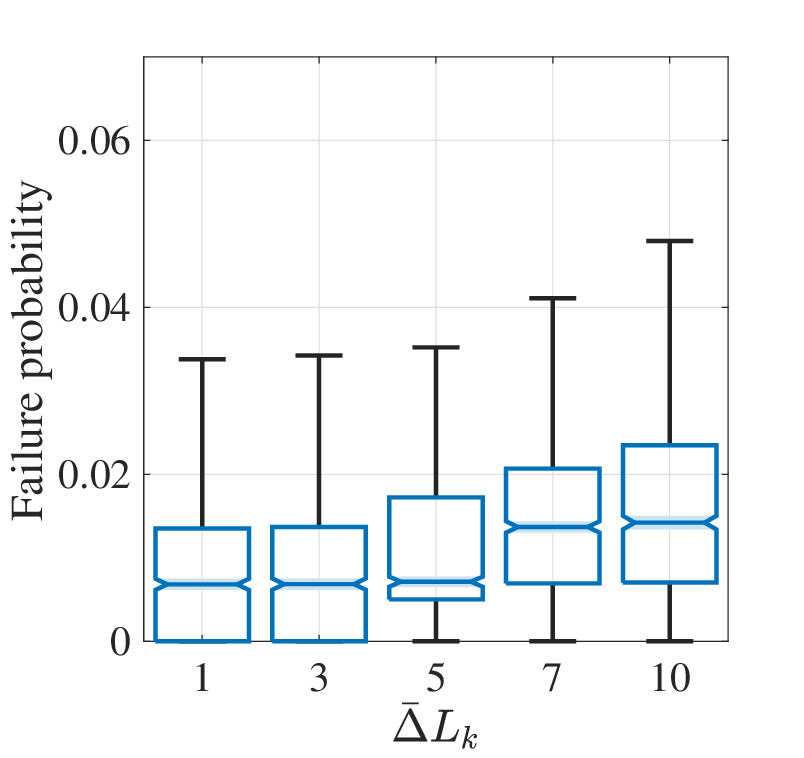} \\ 
		\vspace*{-0pt}
		\centering {\footnotesize$(b)$  Failure probability}
		\vspace*{-0pt}
	\end{minipage}
        \captionsetup{format=plain, justification=justified, width=1\linewidth}
	\caption{Performance evaluation of AoL-REVERB solution under different AoL requirements $\bar{\Delta}L_k$.}
	\label{fig_AoL_CDF_failure}
\vspace*{-15pt}
\end{figure}
Fig.~\ref*{fig_PRBvsMRMSE} illustrates different schemes' PRB consumption and Mean-root-mean-square-error (MRMSE). As expected, AoL-REVERB  proves to be the most efficient, consuming the least resource and achieving as same MRMSE compared to \textit{CB-Greedy} and \textit{EB-Greedy}. On average, AoL-REVERB consumes around $2900$ PRBs to reach the goal, compared to $6200$ PRBs  of the \textit{CB-Greedy} and $6800$ PRBs  of the \textit{EB-Greedy}. Even though \textit{Traditional} one only queries 2 $\SA$s per QI, the power consumption is also as same as AoL-REVERB because this scheme takes more QIs to reach the goal, as illustrated in Fig.~\ref{fig_trajectory}(c). At the same time, AoL-REVERB's MRMSE is maintained at approximately the same low level as the \textit{CB-Greedy} and \textit{EB-Greedy} methods while \textit{Traditional} has traded off low PRB consumption with up to $90\times$ higher in terms of MRMSE comparing to others.
The results achieved in Fig.~\ref*{fig_PRBvsMRMSE} are explained by a snapshot of the uncertainty evolution and the strategic selection of $\SA$s in Fig.~\ref{fig_uncertainty_evolution}, based on their contributions to the DT's performance. It is noteworthy that the management of AoL-REVERB's uncertainty is subject to the control of both DT and RL requirements as flowing \eqref{glob_qos}. The requisites imposed upon the RL agent's behavior are primarily administered through the reward function delineated in \eqref{reward_shaping1}. Hence, DT only requests more observations when the DT threshold is surpassed or when the RL agent necessitates high precision, typically when the agent is nearing its goal and precise force control is imperative.

In Fig.~\ref{fig_vary_Cmax}, we assess the efficacy of the proposed algorithm under varying communication capacity, specifically analyzing the impact of changes in the maximum number of simultaneous connections allowed at each QI. Fig.~\ref{fig_vary_Cmax}(a) illustrates the total PRBs utilized by various schemes. It is evident that the \textit{traditional}  use the same PRB usage even with an increase in $C_{}$, as it consistently demands two observations (position and velocity) at each QI. Notably, AoL-REVERB consistently maintains a relatively constant total PRB usage, averaging around 3300 PRBs, irrespective of the number of maximum connections. Meanwhile, the PRB consumption for systems implementing \textit{CB-Greedy} and \textit{EB-Greedy} is much higher, averaging {9200} and {12800} PRBs, respectively, representing approximately $2.79\times$ and  $3.88\times$ higer than AoL-REVERB at $C=30$. This phenomenon can be elucidated by the fact that our approach only requests $\SA$s until the certainty requirement at DT is met, ensuring optimal control at each QI, as indicated in step 3 of Algorithm 2. Thus, the prior estimation \eqref{mu_prior} is sufficient to attain the requisite confidence level. This becomes apparent when the number of allowed connections is low $(C_{} < 10)$, as depicted in Fig.~\ref{fig_vary_Cmax}(b) and Fig.~\ref{fig_vary_Cmax}(c). In particular, Fig.~\ref{fig_vary_Cmax}(b) presents the cumulative distribution function (CDF), while Fig.~\ref{fig_vary_Cmax}(c) showcases the failure probability concerning the confidence of estimation at DT for various $C$ levels. It is observed that when $C$  ranges from 1 to 10, PRB consumption rapidly increases (averaging from {721} to {2900} in Fig.~\ref{fig_vary_Cmax}(b)) to effectively reduce the failure probability from 0.18 to 0.015, as seen in Fig.~\ref{fig_vary_Cmax}(c). However, as $C$  surpasses 15, PRB consumption plateaus and remains nearly constant at 3300. This coincides with the consistently low failure probability as almost 0 in Fig.~\ref{fig_vary_Cmax}(c).

\phuc{To evaluate the impact of timing on system performance, we emphasize that the required uplink latency directly affects the consumed PRBs. However, the latency requirement within constraint \eqref{glob_problemc} is addressed with a closed-form solution in Section IV(C), resulting in the cost and latency relationship illustrated in equation \eqref{BW_require}. Consequently, in this section, we examine the influence of AoL, which impacts the control performance in considered DT architecture.} Fig.~\ref{fig_AoL_status} illustrates the distribution of AoL, depicting both the position status in Fig.~\ref{fig_AoL_status}(a) and velocity in Fig.~\ref{fig_AoL_status}(b) within the position-velocity coordinate system, under setting $\bar{\Delta}L_k=5,  \forall k\in\Kcal$. The AoL value of the position state typically ranges between 1.5 and 2.5, whereas the velocity state ranges between 2.5 and 3.5. This observation suggests that the position exerts a greater influence on the calculation of the optimal control value, leading to more frequent updates of the position state compared to the velocity state under this configuration.

To specifically exploit the impact of the AoL requirement on system performance, Fig.~\ref{fig_AoL_CDF_failure} presents the CDF of the number of selected SAs and the failure probability across different AoL thresholds. In particular, Fig.\ref{fig_AoL_CDF_failure}(a) illustrates the CDF of the number of selected SAs as $\bar{\Delta}L_k$ varies from 1 to 10. Notably, for AoL requirement $\bar{\Delta}L_k$ of 1, 5, and 10, the average number of selected $\SA$s provided by Algorithm 2 is $5.3$, $3.87$, and $3.74$, respectively. This implies that the $\SA$ request frequency is $1.37 \times$ and $1.41 \times$ lower for $\bar{\Delta}L_k =5 $ and $\bar{\Delta}L_k = 10$, respectively, compared to $\bar{\Delta}L_k=1, \forall k \in \Kcal$. It is apparent that with lower AoL values, the system needs to request $\SA$s more frequently to ensure that condition \eqref{glob_probleme} is met. However, as AoL conditions become less stringent with increasing AoL values (i.e., $\bar{\Delta}L_k \geq 5$), the system still needs to gather observations to fulfill the certainty \eqref{glob_problemd} and control \eqref{eta_compute} requirements, leading to the fact that the number of selected $\SA$s does not decrease linearly with the increase in $\bar{\Delta}L_k$. This is further confirmed by the failure probability in Fig.~\ref{fig_AoL_CDF_failure}(b), which remains relatively constant across low values of AoL thresholds $\bar{\Delta}L_k \leq 5$. This observation suggests that the system's performance is not significantly affected by the AoL requirement, as the system is able to maintain a low failure probability across different AoL thresholds. This is further confirmed by the failure probability depicted in Fig.~\ref{fig_AoL_CDF_failure}(b), which remains relatively constant for low values of AoL thresholds $\bar{\Delta}L_k \leq 5$ but increases to $1.31\times$ and $1.96 \times$ the value observed at $\bar{\Delta}L_k =1$ for $\bar{\Delta}L_k =5$ and $\bar{\Delta}L_k =10$, respectively. This underscores the flexibility of Algorithm 2 in adapting to varying AoL requirements by adjusting the number of selected $\SA$s to fulfill both the uncertainty requirements of DT and the control solution.

\vspace*{-5pt}
%%%%%%%%%%%%%%%%%%%%%%%%%%%%%%%%%%%%%%%%%%%%%%%%
\section{Conclusions}
%%%%%%%%%%%%%%%%%%%%%%%%%%%%%%%%%%%%%%%%%%%%%%%%
This work introduced the DT framework for reliability monitoring, predicting and controlling of a communication system. Under the timing constraint about latency, reliability and Age-of-Loop, the  DT platform was shown to obtain more reliable control and trajectory tracking than conventional methods while significant saving communication cost. This result is achieved thanks to the proposed uncertainty control POMDP scheme combined with an efficient algorithm selecting subsets of partial $\SA$s to meet the requirements in the confidence of state estimation and AoL. Future study could explore the long-term impacts of scheduling decisions in complex systems and incorporate deep learning-based estimators.

\vspace*{-5pt}
\setstretch{0.9}
\bibliographystyle{IEEEtran}
\bibliography{Journal}
%\vspace{-0.2cm}
\end{document}